

\message{Preloading the phys format:}


\catcode`\@=11

\chardef\f@ur=4
\chardef\l@tter=11
\chardef\@ther=12
\dimendef\dimen@iv=4
\toksdef\toks@i=1 
\toksdef\toks@ii=2
\newtoks\emptyt@ks 

\def\glet{\global\let}
\def\gz@#1{\global#1\z@}
\def\gm@ne#1{\global#1\m@ne}
\def\g@ne#1{\global\advance#1\@ne}

\def\@height{height}       
\def\@depth{depth}         
\def\@width{width}         

\def\@plus{plus}           
\def\@minus{minus}         

\message{macros for text,}


\def\loop#1\repeat{\def\iter@te{#1\expandafter\iter@te \fi}\iter@te
  \let\iter@te\undefined}

\def\bpargroup{\bp@rgroup\ep@r} 
\def\bgrafgroup{\bp@rgroup\ep@rgroup} 
\def\bp@rgroup{\bgroup \let\par\ep@rgroup \let\endgraf}

\def\ep@r{\ifhmode \unpenalty\unskip \fi \p@r}
\def\ep@rgroup{\ep@r \egroup}

\let\p@r=\endgraf   
\let\par=\ep@r
\let\endgraf=\ep@r

\def\lb{\hfil\break}
\def\endpage{\par \vfil \eject}
\def\superendpage{\par \vfil \supereject}



\def\leftline{\@line\hsize\empty\hss}
\def\rightline{\@line\hsize\hss\empty}
\def\centerline{\@line\hsize\hss\hss}

\let\plainrlap=\rlap   
\let\plainllap=\llap   

\def\rlap{\@line\z@\empty\hss}
\def\llap{\@line\z@\hss\empty}

\def\lftline{\@line\hsize\empty\hfil}
\def\rtline{\@line\hsize\hfil\empty}
\def\ctrline{\@line\hsize\hfil\hfil}

\def\@line#1#2#3{\hbox to#1\bgroup#2\let\n@xt#3%
  \afterassignment\@@line \setbox\z@\hbox}
\def\@@line{\aftergroup\@@@line}
\def\@@@line{\unhbox\z@ \n@xt\egroup}

\def\after@arg#1{\bgroup\aftergroup#1\afterassignment\after@@arg\@eat}
\def\after@@arg{\ifcat\bgroup\noexpand\n@xt\else \n@xt\egroup \fi}
\def\@eat{\let\n@xt= } 
\def\eat#1{}           
\def\@eat@#1{\@eat}    


\let\nl=\space

\def\ctrlines#1#2{\par \bpargroup
  \bgroup \parskip\z@skip \noindent \egroup
  \let\ctr@style#1\let\nl\ctr@lines \hfil \ctr@style{#2}\strut
  \interlinepenalty\@M \par}
\def\ctr@lines{\strut \lb \strut \hfil \ctr@style}


\def\begin@{\ifmmode \expandafter\mathpalette\expandafter\math@ \else
  \expandafter\make@ \fi}
\def\make@#1{\setbox\z@\hbox{#1}\fin@}
\def\math@#1#2{\setbox\z@\hbox{$\m@th#1{#2}$}\fin@}

\def\ph@nt{\let\fin@\finph@nt \begin@}
\let\makeph@nt=\undefined
\let\mathph@nt=\undefined

\newif\ift@ \newif\ifb@
\def\topsmash{\t@true\b@false\sm@sh}
\def\botsmash{\t@false\b@true\sm@sh}
\def\smash{\t@true\b@true\sm@sh}
\def\sm@sh{\let\fin@\finsm@sh \begin@}
\let\makesm@sh=\undefined
\let\mathsm@sh=\undefined
\def\finsm@sh{\ift@\ht\z@\z@\fi \ifb@\dp\z@\z@\fi \box\z@}


\newdimen\boxitsep   \boxitsep=5pt

\def\fboxit#1#2{\hbox{\vrule \@width#1\p@
    \vtop{\vbox{\hrule \@height#1\p@ \vskip\boxitsep
        \hbox{\hskip\boxitsep #2\hskip\boxitsep}}%
      \vskip\boxitsep \hrule \@height#1\p@}\vrule \@width#1\p@}}


\begingroup
  \catcode`\:=\active
  \lccode`\*=`\\ \lowercase{\gdef:{*}}   
  \catcode`\;=\active
  \lccode`\* `\% \lowercase{\gdef;{*}}   
  \catcode`\^^M=\active \glet^^M=\space  
\endgroup


\begingroup
  \catcode`\:=\active
  \outer\gdef\comment{\begingroup
    \catcode`\\\@ther \catcode`\%\@ther \catcode`\^^M\@ther
    \catcode`\{\@ther \catcode`\}\@ther \catcode`\#\@ther
    \wlog{* input between `:comment' and `:endcomment' ignored *}%
    \c@mment}
\endgroup
{\lccode`\:=`\\ \lccode`\;=`\^^M
  \lowercase{\gdef\c@mment#1;{\c@@mment:endcomment*}}}
\def\c@@mment#1#2*#3{\if #1#3%
    \ifx @#2@\def\n@xt{\endgroup\ignorespaces}\else
      \def\n@xt{\c@@mment#2*}\fi \else
    \def\n@xt{\c@mment#3}\fi \n@xt}

\message{date and time,}


\newcount\langu@ge

\let\mainlanguage\relax

\begingroup \catcode`\"=\@ther \gdef\dq{"}
  \catcode`\"=\active
  \gdef"#1{\ifx#1s\ss\else\ifx#1SSS\else
    {\accent\dq 7F #1}\penalty\@M \hskip\z@skip \fi\fi}
  \endgroup

\outer\def\english{\gm@ne\langu@ge
  \global\catcode`\"\@ther \glet\3\undefined}
\outer\def\german{\gz@\langu@ge
  \global\catcode`\"\active \glet\3\ss}

\def\case@language#1{\ifcase\expandafter\langu@ge #1\fi}
\def\case@abbr#1{{\let\nodot\n@dot\case@language{#1}.~}}
\def\n@dot{\expandafter\eat}
\let\nodot=\empty


\def\themonth{\xdef\themonth{\noexpand\case@language
  {\ifcase\month \or Januar\or Februar\or M\noexpand\"arz\or April\or
  Mai\or Juni\or Juli\or August\or September\or Oktober\or November\or
  Dezember\fi
  \noexpand\else
  \ifcase\month \or January\or February\or March\or April\or May\or
  June\or July\or August\or September\or October\or November\or
  December\fi}}\themonth}

\def\thedate{\case@language{\else\themonth\ }\number\day
  \case@language{.\ \themonth \else ,} \number\year}

\def\date{\number\day.\,\number\month.\,\number\year}


\def\PhysTeX{$\Phi\kern-.25em\raise.4ex\hbox{$\Upsilon$}\kern-.225em
  \Sigma$-\TeX}

\message{the time,}



\count255=\time \divide\count255 by 60 \edef\thetime{\the\count255 :}
\multiply\count255 by -60 \advance\count255 by\time
\edef\thetime{\thetime \ifnum10>\count255 0\fi \the\count255 }

\message{spacing, fonts and sizes,}


\newskip\refbetweenskip   \newskip\chskiptamount
\newskip\chskiplamount   \newskip\secskipamount
\newskip\footnotebaselineskip   \newskip\interfootnoteskip

\newdimen\chapstretch   \chapstretch=2.5cm
\newcount\chappenalty   \chappenalty=-800
\newdimen\sectstretch   \sectstretch=2cm
\newcount\sectpenalty   \sectpenalty=-400

\def\chskipt{\chapbreak \vskip\chskiptamount}
\def\chskipl{\nobreak \vskip\chskiplamount}
\def\unchskip{\vskip-\chskiplamount}
\def\secskipt{\sectbreak \vskip\secskipamount}
\def\chapbreak{\par \vskip\z@\@plus\chapstretch \penalty\chappenalty
  \vskip\z@\@plus-\chapstretch}
\def\sectbreak{\par \vskip\z@\@plus\sectstretch \penalty\sectpenalty
  \vskip\z@\@plus-\sectstretch}



\begingroup \lccode`\*=`\r
  \lowercase{\def\n@xt#1*#2@{#1}
    \xdef\font@sel{\expandafter\n@xt\fontname\tenrm*@}}\endgroup

\message{loading \font@sel\space fonts,}

=\font@sel r12 
=\font@sel r9
=\font@sel r8
=\font@sel r6

=\font@sel mi12 \skewchar\twelvei='177 
=\font@sel mi9    \skewchar\ninei='177
=\font@sel mi8   \skewchar\eighti='177
=\font@sel mi6   \skewchar\sixi='177

=\font@sel sy10 scaled \magstep1
  \skewchar\twelvesy='60 
=\font@sel sy9    \skewchar\ninesy='60
=\font@sel sy8   \skewchar\eightsy='60
=\font@sel sy6   \skewchar\sixsy='60




=\font@sel bx12 
=\font@sel bx9
=\font@sel bx8
=\font@sel bx6

=\font@sel tt12 
=\font@sel tt8


=\font@sel sl12 
=\font@sel sl9
=\font@sel sl8

\font\twelveit=\font@sel ti12 
\font\nineit=\font@sel ti9
\font\eightit=\font@sel ti8
=\font@sel ti7



=\font@sel csc10 scaled \magstep1 
=\font@sel csc10






\def\twelvepoint{\twelve@point
  \let\normal@spacing\twelve@spacing \set@spacing}
\def\tenpoint{\ten@point
  \let\normal@spacing\ten@spacing \set@spacing}
\def\eightpoint{\eight@point
  \let\normal@spacing\eight@spacing \set@spacing}

\def\rm{\fam\z@ \@fam}
\def\mit{\fam\@ne}
\def\oldstyle{\mit \@fam}
\def\cal{\fam\tw@}
\def\it{\fam\itfam \@fam}
\def\sl{\fam\slfam \@fam}
\def\bf{\fam\bffam \@fam}
\def\tt{\fam\ttfam \@fam}
\def\caps{\@caps}
\def\@fam{\the\textfont\fam}


\def\twelve@point{\set@fonts twelve ten eight }
\def\ten@point{\set@fonts ten eight six }
\def\eight@point{\set@fonts eight six five }

\def\set@fonts#1 #2 #3 {\textfont\ttfam\csname#1tt\endcsname
    \expandafter\let\expandafter\@caps\csname#1csc\endcsname
  \def\n@xt##1##2{\textfont##1\csname#1##2\endcsname
    \scriptfont##1\csname#2##2\endcsname
    \scriptscriptfont##1\csname#3##2\endcsname}%
  \set@@fonts}
\def\set@@fonts{\n@xt0{rm}\n@xt1i\n@xt2{sy}%
  \textfont3\tenex \scriptfont3\tenex \scriptscriptfont3\tenex
  \n@xt\itfam{it}\n@xt\slfam{sl}\n@xt\bffam{bf}\rm}


\def\singlespace{\chardef\@spacing\z@ \set@spacing}
\def\doublespace{\chardef\@spacing\@ne \set@spacing}
\def\triplespace{\chardef\@spacing\tw@ \set@spacing}
\chardef\@spacing=1   

\def\set@spacing{\expandafter\expandafter\expandafter\set@@spacing
  \expandafter\spacing@names\expandafter\@@\normal@spacing
  \normalbaselines}
\def\set@@spacing#1#2\@@#3+#4*{#1#4\multiply#1\@spacing \advance#1#3%
  \ifx @#2@\let\n@xt\empty \else
    \def\n@xt{\set@@spacing#2\@@}\fi \n@xt}

\def\normalbaselines{\lineskip\normallineskip
  \setbaselineskip\normalbaselineskip
  \lineskiplimit\normallineskiplimit}

\def\setbaselineskip{\afterassignment\set@strut \baselineskip}
\def\set@strut{\setbox\strutbox\spacer\z@\baselineskip}

\def\spacer{\hbox\bgroup \afterassignment\x@spacer \dimen@}
\def\x@spacer{\ifdim\dimen@=\z@\else \hskip\dimen@ \fi
  \afterassignment\y@spacer \dimen@}
\def\y@spacer{\setbox\z@\hbox{$\vcenter{\vskip\dimen@}$}%
  \vrule \@height\ht\z@ \@depth\dp\z@ \@width\z@ \egroup}

\def\spacing@names{
  \normalbaselineskip
  \normallineskip
  \normallineskiplimit
  \footnotebaselineskip
  \interfootnoteskip
  \parskip
  \refbetweenskip
  \abovedisplayskip
  \belowdisplayskip
  \abovedisplayshortskip
  \belowdisplayshortskip
  \chskiptamount
  \chskiplamount
  \secskipamount
  }

\def\twelve@spacing{
  14\p@              +5\p@        *
  \p@                +\z@         *
  \z@                +\z@         *
  14\p@              +\p@         *
  20\p@              +\z@         *
  5\p@\@plus\p@      +-2\p@       *
  \z@                +4\p@        *
  8\p@\@plus2\p@\@minus3\p@  +%
    4\p@\@plus3\p@\@minus5\p@     *
  8\p@\@plus2\p@\@minus3\p@  +%
    4\p@\@plus3\p@\@minus5\p@     *
  \p@\@plus2\p@\@minus\p@    +%
    4\p@\@plus3\p@\@minus2\p@     *
  8\p@\@plus2\p@\@minus3\p@  +%
    \p@\@plus2\p@\@minus2\p@      *
  20\p@\@plus5\p@    +\z@         *
  5.5\p@             +\z@         *
  6\p@\@plus2\p@     +\z@         *
  }

\def\ten@spacing{
  11\p@              +4.5\p@      *
  \p@                +\z@         *
  \z@                +\z@         *
  12\p@              +\p@         *
  16\p@              +\z@         *
  5\p@\@plus\p@      +-2\p@       *
  \z@                +5\p@        *
  8\p@\@plus2\p@\@minus3\p@  +%
    4\p@\@plus3\p@\@minus5\p@     *
  8\p@\@plus2\p@\@minus3\p@  +%
    4\p@\@plus3\p@\@minus5\p@     *
  \p@\@plus2\p@\@minus\p@    +%
    4\p@\@plus3\p@\@minus2\p@     *
  8\p@\@plus2\p@\@minus3\p@  +%
    \p@\@plus2\p@\@minus2\p@      *
  20\p@\@plus5\p@    +\z@         *
  5.5\p@             +\z@         *
  6\p@\@plus2\p@     +\z@         *
  }

\def\eight@spacing{
  9\p@               +3.5\p@      *
  \p@                +\z@         *
  \z@                +\z@         *
  10\p@              +\p@         *
  14\p@              +\z@         *
  5\p@\@plus\p@      +-2\p@       *
  \z@                +5\p@        *
  8\p@\@plus2\p@\@minus3\p@  +%
    4\p@\@plus3\p@\@minus5\p@     *
  8\p@\@plus2\p@\@minus3\p@  +%
    4\p@\@plus3\p@\@minus5\p@     *
  \p@\@plus2\p@\@minus\p@    +%
    4\p@\@plus3\p@\@minus2\p@     *
  8\p@\@plus2\p@\@minus3\p@  +%
    \p@\@plus2\p@\@minus2\p@      *
  20\p@\@plus5\p@    +\z@         *
  5.5\p@             +\z@         *
  6\p@\@plus2\p@     +\z@         *
  }

\twelvepoint


\def\large{\par \bgroup \twelvepoint \after@arg\@size}
\def\medium{\par \bgroup \tenpoint \after@arg\@size}
\def\small{\par \bgroup \eightpoint \after@arg\@size}
\def\@size{\par \egroup}

\def\LARGE#1{{\twelve@point #1}}
\def\MEDIUM#1{{\ten@point #1}}
\def\SMALL#1{{\eight@point #1}}

\message{texts, headings and styles,}


\def\submittextone{Zur Ver\"offentlichung in\else Submitted to}
\def\submittexttwo{ eingereicht\else}
\def\abstracthead{Zusammenfassung\else Abstract}
\def\ackhead{Danksagung\else Acknowledgements}
\def\appendixhead{Anhang\else Appendix}
\def\eqabbr{Gl\else eq}
\def\eqsabbr{Gln\else eqs}
\def\figpref{Abb\else Fig}
\def\fighead{Abbildungen\else Figure captions}
\def\figabbr{Bild\nodot\else Fig}
\def\figsabbr{Bilder\nodot\else Figs}
\def\tabpref{Tab\else Tab}
\def\tabhead{Tabellen\else Table captions}
\let\tababbr=\tabpref
\def\tabsabbr{Tab\else Tabs}
\def\refpref{Lit\else Ref}
\def\refhead{Literaturverzeichnis\else References}
\def\refabbr{???\else Ref}
\def\refsabbr{????\else Refs}
\def\tocpref{Inh\else Toc}
\def\tochead{Inhaltsverzeichnis\else Table of contents}
\def\footpref{Anm\else Foot}
\def\foothead{Anmerkungen\else Footnotes}
\def\prfhead{Beweis\else Proof}


\def\UPPERCASE#1{\edef\n@xt{#1}\uppercase\expandafter{\n@xt}}

\let\headlinestyle=\twelverm
\let\footlinestyle=\twelverm
\let\pagestyle=\twelverm       
\let\titlestyle=\bf            
\let\authorstyle=\caps
\let\addressstyle=\sl
\let\sectstyle=\caps           

\let\headstyle=\UPPERCASE      
\let\captionstyle=\it          
\let\journalstyle=\sl
\let\volumestyle=\bf



\def\namrefindent{2em}


\let\footstyle=\empty          

\let\stmttitlestyle=\bf        
\let\stmtstyle=\sl
\let\prftitlestyle=\caps
\let\prfstyle=\sl


\def\skipuserexit{\setbox\z@\box\@cclv}  
\def\shipuserexit{\unvbox\@cclv}         

\def\chapuserexit{\sectuserexit}         
\def\appuserexit{\chapuserexit}          
\def\sectuserexit{\secsuserexit}         
\let\secsuserexit=\relax                 

\message{page numbers and output,}


\newcount\firstp@ge   \firstp@ge=-10000
\newcount\lastp@ge   \lastp@ge=10000
\outer\def\pagesel#1#2{\global\firstp@ge#1 \global\lastp@ge#2
  \wlog{**************************}\wlog{*}%
  \wlog{* don't use \string\p agesel}%
  \wlog{* this will not be supported in future}%
  \wlog{*}\wlog{**************************}%
  \wlog{(* pages #1-#2 selected for printing, others will be skipped *)}}

\newbox\pageb@x

\outer\def\toppagenum{\glet\page@tbn T%
  \glet\headb@x\pageb@x \glet\footb@x\voidb@x}
\outer\def\botpagenum{\glet\page@tbn B%
  \glet\headb@x\voidb@x \glet\footb@x\pageb@x}
\outer\def\nopagenum{\glet\page@tbn N%
  \glet\headb@x\voidb@x \glet\footb@x\voidb@x}

\outer\def\lefthead{\glet\head@lrac L}
\outer\def\righthead{\glet\head@lrac R}
\outer\def\althead{\glet\head@lrac A}
\outer\def\centhead{\glet\head@lrac C}
\outer\def\leftfoot{\glet\foot@lrac L}
\outer\def\rightfoot{\glet\foot@lrac R}
\outer\def\altfoot{\glet\foot@lrac A}
\outer\def\centfoot{\glet\foot@lrac C}

\newtoks\lheadtext   \newtoks\cheadtext   \newtoks\rheadtext
\newtoks\lfoottext   \newtoks\cfoottext   \newtoks\rfoottext

\headline={\headlinestyle \head@foot\skip@head\head@lrac
  \lheadtext\cheadtext\rheadtext\headb@x}
\footline={\footlinestyle \head@foot\skip@foot\foot@lrac
  \lfoottext\cfoottext\rfoottext\footb@x}
\lheadtext={}   \cheadtext={}   \rheadtext={}
\lfoottext={}   \cfoottext={}   \rfoottext={}

\newbox\page@strut
\setbox\page@strut\hbox{\vrule \@height 15mm\@depth 10mm\@width \z@}

\def\head@foot#1#2#3#4#5#6{\unhcopy\page@strut
  \if#1T\hfil \else
    \if#2C\head@@foot{\the#3}{\copy#6}{\the#5}\else
      \if#2A\ifodd\pageno \let#2R\else \let#2L\fi \fi
      \if#2R\head@@foot{\the#4}{\the#5}{\copy#6}\else
        \head@@foot{\copy#6}{\the#3}{\the#4}\fi \fi \fi}
\def\head@@foot#1#2#3{\plainrlap{#1}\hfil#2\hfil\plainllap{#3}}

\let\startpage=\relax  

\outer\def\pageall{\glet\page@ac A%
  \global\countdef\pageno\z@ \global\pageno\@ne
  \global\countdef\pageno@pref\@ne \gz@\pageno@pref
  \glet\page@pref\empty \glet\page@reset\count@
  \glet\chap@break\chskipt \outer\gdef\startpage{\global\pageno}}
\outer\def\pagechap{\glet\page@ac C%
  \global\countdef\pageno\@ne \gz@\pageno
  \global\countdef\pageno@pref\z@ \gz@\pageno@pref
  \gdef\page@pref{\dash@pref}%
  \gdef\page@reset{\global\pageno\@ne \global\pageno@pref}%
  \glet\chap@break\superendpage
  \outer\gdef\startpage##1.{\global\pageno@pref##1\global\pageno}}


\hsize=15 cm   \hoffset=0 mm
\vsize=22 cm   \voffset=0 mm

\newdimen\hoffset@corr@p   \newdimen\voffset@corr@p
\newdimen\hoffset@corrm@p   \newdimen\voffset@corrm@p
\newdimen\hoffset@corr@l   \newdimen\voffset@corr@l
\newdimen\hoffset@corrm@l   \newdimen\voffset@corrm@l

\outer\def\portrait{\switch@pl P%
  \glet\hoffset@corr\hoffset@corr@p
  \glet\voffset@corr\voffset@corr@p
  \glet\hoffset@corrm\hoffset@corrm@p
  \glet\voffset@corrm\voffset@corrm@p}
\outer\def\landscape{\switch@pl L%
  \glet\hoffset@corr\hoffset@corr@l
  \glet\voffset@corr\voffset@corr@l
  \glet\hoffset@corrm\hoffset@corrm@l
  \glet\voffset@corrm\voffset@corrm@l}
\def\switch@pl#1{\if #1\ori@pl \else \superendpage \glet\ori@pl#1%
  \dimen@\ht\page@strut \advance\dimen@\dp\page@strut
  \advance\vsize\dimen@ \dimen@ii\hsize \global\hsize\vsize
  \advance\dimen@ii-\dimen@ \global\vsize\dimen@ii \fi}
\let\ori@pl=P

\def\m@g{\dimen@\ht\page@strut \advance\dimen@\dp\page@strut
  \advance\vsize\dimen@ \divide\vsize\count@
  \multiply\vsize\mag \advance\vsize-\dimen@
  \divide\hsize\count@ \multiply\hsize\mag
  \divide\dimen\footins\count@ \multiply\dimen\footins\mag
  \mag\count@}

\output={\physoutput}

\def\physoutput{\make@lbl
  \ifnum \pageno<\firstp@ge \skipp@ge \else
  \ifnum \pageno>\lastp@ge \skipp@ge \else \shipp@ge \fi \fi
  \advancepageno \skippagenum F\skipheadline F\skipfootline F%
  \ifnum\outputpenalty>-\@MM \else \dosupereject \fi}

\def\skippagenum{\glet\skip@page}
\def\skipheadline{\glet\skip@head}
\def\skipfootline{\glet\skip@foot}

\def\skipp@ge{{\skipuserexit \setbox\z@\box\topins
  \setbox\z@\box\footins}\deadcycles\z@}
\def\shipp@ge{\setbox\pageb@x\hbox{%
    \if F\skip@page \pagestyle{\page@pref \folio}\fi}%
  \dimen@-.5\hsize \advance\dimen@\hoffset@corrm
  \divide\dimen@\@m \multiply\dimen@\mag
  \advance\hoffset\dimen@ \advance\hoffset\hoffset@corr
  \dimen@\ht\page@strut \advance\dimen@\dp\page@strut
  \advance\dimen@\vsize \dimen@-.5\dimen@
  \advance\dimen@\voffset@corrm
  \divide\dimen@\@m \multiply\dimen@\mag
  \advance\voffset\dimen@ \advance\voffset\voffset@corr
  \shipout\vbox{\makeheadline \vbadness\@M \setbox\z@\pagebody
    \dimen@\dp\z@ \box\z@ \kern-\dimen@ \makefootline}}

\def\pagecontents{\ifvbox\topins\unvbox\topins\fi
  \dimen@\dp\@cclv \shipuserexit 
  \ifvbox\footins 
    \vskip\skip\footins \footnoterule \unvbox\footins\fi
  \ifr@ggedbottom \kern-\dimen@ \vfil \fi}

\def\folio{\ifnum\pageno<\z@ \ifcase\langu@ge \MEDIUM{\uppercase
  \expandafter{\romannumeral-\pageno}}\else \romannumeral-\pageno \fi
  \else \number\pageno \fi}

\def\makeheadline{\line{\the\headline}\nointerlineskip}
\def\makefootline{\nointerlineskip \line{\the\footline}}

\skippagenum=F   \skipheadline=F   \skipfootline=F

\message{title page macros,}


\outer\def\titlepage{\glet\titl@fill\vfil}
\outer\def\notitlepage{\gdef\titl@fill{\vskip20\p@}}

\newbox\t@pleft   \newbox\t@pright
\def\t@pinit{%
  \global\setbox\t@pleft\vbox{\hrule \@height\z@ \@width.26\hsize}%
  \global\setbox\t@pright\copy\t@pleft}
\t@pinit

\def\topleft{\t@p\t@pleft}
\def\topright{\t@p\t@pright}
\def\t@p#1#2{\global\setbox#1\vtop{\unvbox#1\hbox{\strut #2}}}

\outer\def\submit#1{\topleft{\case@language\submittextone}%
  \topleft{{#1}\case@language\submittexttwo}}

\let\pubdate=\topright

\outer\def\title{\vbox{\line{\box\t@pleft \hss \box\t@pright}}%
  \skippagenum T\skipheadline T\skipfootline T%
  \t@pinit \titl@fill \vskip\chskiptamount \@title}
\let\titcon=\relax  
\outer\def\titcon{\errmessage{please use \noexpand\nl in the title
  instead of \noexpand\titcon}\unchskip \@title}
\def\@title#1{\ctrlines\titlestyle{#1}\chskipl}
\def\titl@#1{\edef\n@xt{\noexpand\@title{#1}}\n@xt}

\def\author{\aut@add\authorstyle}
\def\autcon{\and@con \author}
\def\address{\aut@add\addressstyle}
\def\addcon{\and@con \address}
\def\and@con{\titl@fill \ctrline{\case@language{und\else and}}}

\def\aut@add#1{\titl@fill \ctrlines{#1\use@nl}}
\def\use@nl{\let\\\use@@nl}
\def\use@@nl{\errmessage{please use \noexpand\nl in addresses and
  (lists of) authors instead of \string\\}\nl}

\def\abstract{\titl@fill \he@d{\case@language\abstracthead}%
  \after@arg\titl@fill}

\def\ack{\chskipt \he@d{\case@language\ackhead}}

\def\he@d#1{\ctrline{\headstyle{#1}}\chskipl}

\message{chapters, sections and appendices,}


\newtoks\l@names   \l@names={\\\the@label}
\let\the@label=\empty

\def\label{\num@lett\@label}
\def\@label#1{\def@name\l@names#1{\the@label}}
\def\quote{\num@lett\empty}

\newinsert\lbl@ins
\count\lbl@ins=0   \dimen\lbl@ins=\maxdimen   \skip\lbl@ins=0pt
\newcount\lbln@m   \lbln@m=0
\let\lbl@saved\empty

\def\pagelabel{\num@lett\@pagelabel}
\def\@pagelabel#1{\ifx#1\undefined \let#1\empty \fi
  \toks@\expandafter{#1}\expandafter\testcr@ss\the\toks@\cr@ss\@@
  \ifcr@ss\else
    \toks@{\cr@ss\lbl@undef}\def@name\l@names#1{\the\toks@}\fi
  \g@ne\lbln@m \insert\lbl@ins{\vbox{\vskip\the\lbln@m sp}}%
  \count@\lbln@m \do@label\store@label#1}
\begingroup \let\save=\relax  
  \gdef\lbl@undef{\message{unresolved \string\pagelabel, use
    \string\save\space and \string\crossrestore}??}
\endgroup
\def\do@label#1{\expandafter#1\csname\the\count@\endcsname}
\def\store@label#1#2{\expandafter\gdef\expandafter\lbl@saved
  \expandafter{\lbl@saved#1#2}}

\def\make@lbl{\setbox\z@\vbox{\let\MEDIUM\relax
  \unvbox\lbl@ins \loop \setbox\z@\lastbox \ifvbox\z@
    \count@\ht\z@ \do@label\make@label \repeat}}
\def\make@label#1{\def\make@@label##1#1##2##3#1##4\@@{%
    \gdef\lbl@saved{##1##3}%
    \ifx @##4@\errmessage{This can't happen}\else
    \def@name\l@names##2{\page@pref\folio}\fi}%
  \expandafter\make@@label\lbl@saved#1#1#1\@@}

\outer\def\lblrestore{\all@restore\l@names}


\let\sect=\relax  \let\s@ct=\relax  

\def\chapinit{\chap@init{\chap@pref}\glet\sect@@eq\@chap@sect@eq
  \sectinit}
\def\appinit{\chap@init{\char\the\appn@m}\glet\sect@@eq\@chap@eq
  \glet\sect@dot@pref\empty \glet\sect@pref\dot@pref
  \glet\sect\undefined}
\def\sectinit{\xdef\sect@dot@pref{\the\sectn@m.}%
  \xdef\sect@pref{\dot@pref\sect@dot@pref}\glet\sect\s@ct}
\def\chap@init#1{\xdef\the@label{#1}\xdef\dot@pref{\the@label.}%
  \xdef\dash@pref{\the@label--}\glet\chap@@eq\@chap@eq}


\newcount\chapn@m   \chapn@m=0

\outer\def\chappage{\glet\chap@page T}
\outer\def\nochappage{\glet\chap@page F}

\outer\def\arabicchapnum{\glet\chap@ar A\gdef\chap@pref{\the\chapn@m}}
\outer\def\romanchapnum{\glet\chap@ar R%
  \gdef\chap@pref{\uppercase{\romannumeral\chapn@m}}}

\let\chap=\relax  \let\ch@p=\relax  

\outer\def\chapters{\glet\chap@yn Y\glet\chap\ch@p
  \chap@init{0}\glet\sect@@eq\@chap@sect@eq \sectinit}
\outer\def\nochapters{\glet\chap@yn N\glet\chap\undefined
  \glet\dot@pref\empty \glet\dash@pref\empty
  \glet\chap@@eq\@eq \glet\sect@@eq\@sect@eq \sectinit}

\outer\def\ch@p#1{\if T\chap@page \superendpage \else \chap@break \fi
  \g@ne\chapn@m \sect@reset \chapinit \page@reset\chapn@m
  \eq@reset \fig@reset \tab@reset
  \toks@{\dot@pref}\toks@ii{#1}\chapuserexit
  \titl@{\the\toks@\ \the\toks@ii}%
  \ifnum\auto@toc>\m@ne \toks@store{#1}\@toc\dot@pref \fi}



\def\sec@title#1#2#3#4#5#6#{\ifx @#6@\g@ne#1\else\global#1#6\fi
  #2\secskipt \xdef\the@label{#3\the#1}\xdef#4{\the@label.}%
  \read@store{\sec@@title#4#5}}
\def\sec@@title#1#2#3#4{\toks@{\the@label.}\toks@ii\toks@store
  #4\bpargroup #2\varitem{\the\toks@}\interlinepenalty\@M
    \let\nl\lb \the\toks@ii \par\nobreak
  \ifnum#3<\auto@toc \@toc{#1}\fi}

\newcount\sectn@m   \sectn@m=0
\def\sect@reset{\gz@\sectn@m}
\outer\def\s@ct{\sec@title\sectn@m
  {\secs@reset \sectinit \eq@@reset \fig@@reset \tab@@reset}%
  \dot@pref\sect@pref{\sectstyle\z@\sectuserexit}}

\let\secs@reset=\relax


\def\sect@lev{\@ne}         
\def\sect@id{sect}          
\def\secs@id{secs}          

\outer\def\newsect{\begingroup \count@\sect@lev
  \let\@\endcsname \let\or\relax
  \edef\n@xt{\new@sect}\advance\count@\@ne
  \xdef\sect@lev{\the\count@\space}\n@xt
  \glet\sect@id\secs@id \xdef\secs@id{\secs@id s}\endgroup}
\begingroup \let\newcount=\relax
  \gdef\new@sect{\wlog{\noexpand\string\secs@nm\@= subsection
      level \noexpand\sect@lev}\noexpand\newcount\secs@nm\n@m@
    \gdef\secs@nm\@reset@{\noexpand\gz@\secs@nm\n@m@}%
    \outer\gdef\secs@nm\@{\noexpand\sec@title\secs@nm\n@m@
      \secs@nm s\@reset@ \csn@me\sect@id\@pref@ \secs@nm\@pref@
      {\secs@nm\style@{\the\count@}\secs@nm\userexit@}}%
    \glet\secs@nm s\@reset@ \relax
    \gdef\secs@nm\style@{\csn@me\sect@id\style@}%
    \gdef\secs@nm\userexit@{\secs@nm s\userexit@}%
    \glet\secs@nm s\userexit@\relax
    \outer\xdef\csn@me toc\secs@id\@
      {\global\auto@toc\noexpand\the\count@\space}%
    \xdef\noexpand\save@@toc{\save@@toc\or\secs@id}}
\endgroup

\def\n@m@{n@m\@}
\def\@pref@{@pref\@}
\def\@reset@{@reset\@}
\def\style@{style\@}
\def\userexit@{userexit\@}

\def\secs@nm{\csn@me\secs@id}
\def\csn@me{\expandafter\noexpand\csname}


\newcount\appn@m   \appn@m=64

\outer\def\appendix{\if T\chap@page \superendpage \else\chap@break \fi
  \g@ne\appn@m \secs@reset \appinit \page@reset\appn@m
  \eq@reset \fig@reset \tab@reset \futurelet\n@xt \app@ndix}

\def\app@ndix{\ifcat\bgroup\noexpand\n@xt \expandafter\@ppendix \else
  \expandafter\@ppendix\expandafter\unskip \fi}

\def\@ppendix#1{\toks@{\case@language\appendixhead~\dot@pref}
  \toks@ii{#1}\appuserexit
  \titl@{\the\toks@\ \the\toks@ii}%
  \ifnum\auto@toc>\m@ne \toks@store{#1}%
  \@toc{\case@language\appendixhead\ \dot@pref}\fi}

\def\app#1{\chskipt \he@d{\case@language\appendixhead\ #1}}

\message{equations,}


\def\num@lett{\cat@lett \num@@lett}
\def\num@@lett#1#2{\egroup #1{#2}}

\def\num@l@tt{\cat@lett \num@@l@tt}
\def\num@@l@tt#1#{\egroup #1}

\def\cat@lett{\bgroup
  \catcode`\0\l@tter \catcode`\1\l@tter \catcode`\2\l@tter
  \catcode`\3\l@tter \catcode`\4\l@tter \catcode`\5\l@tter
  \catcode`\6\l@tter \catcode`\7\l@tter \catcode`\8\l@tter
  \catcode`\9\l@tter \catcode`\'\l@tter}

\def\quote@all#1{\leavevmode\hbox{\mathcode`\-\dq 707B$#1$}}
\def\use{\num@lett\@use}
\def\@use{\setbox\z@\hbox}


\newtoks\e@names   \e@names={}
\newcount\eqn@m   \eqn@m=0

\outer\def\equall{\glet\eq@acs A\glet\eq@pref\empty
  \glet\def@eq\@eq \glet\eq@reset\relax \glet\eq@@reset\relax}
\outer\def\equchap{\glet\eq@acs C\gdef\eq@pref{\dot@pref}%
  \gdef\def@eq{\chap@@eq}\glet\eq@reset\eqz@ \glet\eq@@reset\relax}
\outer\def\equsect{\glet\eq@acs S\gdef\eq@pref{\sect@pref}%
  \gdef\def@eq{\sect@@eq}\glet\eq@reset\eqz@ \glet\eq@@reset\eqz@}
\def\eqz@{\gz@\eqn@m}

\outer\def\equfull{\glet\eq@fs F\gdef\eq@@fs{\let\test@eq\full@eq}}
\outer\def\equshort{\glet\eq@fs S\glet\eq@@fs\relax}

\def\@eq(#1){#1}
\def\@chap@eq{\noexpand\chap@eq\@eq}
\def\@sect@eq{\noexpand\sect@eq\@eq}
\def\@chap@sect@eq{\noexpand\chap@sect@eq\@eq}

\def\chap@eq{\test@eq\empty\dot@pref}
\def\sect@eq{\test@eq\empty\sect@dot@pref}
\def\chap@sect@eq{\test@eq\sect@eq\dot@pref}
\def\test@eq#1#2#3.{\def\n@xt{#3.}\ifx#2\n@xt \let\n@xt#1\fi \n@xt}
\def\full@eq#1#2{}
\def\short@eq#1#2#3.{#1}

\outer\def\equleft{\glet\eq@lrn L\glet\eqtag\leqno
  \glet\eq@tag\leq@no}
\outer\def\equright{\glet\eq@lrn R\glet\eqtag\eqno
  \glet\eq@tag\eq@no}
\outer\def\equnone{\glet\eq@lrn N\glet\eqtag\n@eqno
  \glet\eq@tag\neq@no}

\begingroup
  \catcode`\$=\active \catcode`\*=3 \lccode`\*=`\$
  \lowercase{\gdef\n@eqno{\catcode`\$\active
                \def$${\egroup **}\setbox\z@\hbox\bgroup}}
\endgroup
\def\eq@no{\llap{$\@lign##$}\tabskip\z@skip}
\def\leq@no{\kern-\displaywidth \rlap{$\@lign##$}\tabskip\displaywidth}
\def\neq@no{\@use{$\@lign##$}\tabskip\z@skip}

\def\displaylines{\afterassignment\display@lines \@eat}
\def\display@lines{\displ@y
  \halign\n@xt\hbox to\displaywidth{$\@lign\hfil\displaystyle##\hfil$}%
    &\span\eq@tag\crcr}

\def\eqalignno{\let\eq@@tag\eq@no \eqalign@tag}
\def\leqalignno{\let\eq@@tag\leq@no \eqalign@tag}
\def\eqaligntag{\let\eq@@tag\eq@tag \eqalign@tag}
\def\eqalign@tag{\afterassignment\eqalign@@tag \@eat}
\def\eqalign@@tag{\displ@y
  \tabskip\centering \halign to\displaywidth\n@xt
    \hfil$\@lign\displaystyle{##}$\tabskip\z@skip
    &$\@lign\displaystyle{{}##}$\hfil\tabskip\centering
    &\span\eq@@tag\crcr}

\def\fulltag#1{{\let\test@eq\full@eq#1}}
\def\shorttag#1{{\let\test@eq\short@eq#1}}

\def\eq{\g@ne\eqn@m \make@eq\empty}
\def\make@eq#1{(\eq@pref\the\eqn@m #1)}
\def\EQ{\eq \num@lett\eq@save}
\def\eq@save#1{\def@name\e@names#1{\expandafter\def@eq\make@eq\empty}}

\def\eqn{\eqtag\eq}
\def\EQN{\eqtag\EQ}

\def\eqadv{\g@ne\eqn@m}
\def\EQADV{\eqadv \num@lett\eq@save}

\newcount\seqn@m   \seqn@m=96

\def\subeqbegin{\global\seqn@m96 \subeq}
\def\SUBEQBEGIN{\global\seqn@m96 \SUBEQ}
\def\subeq{\g@ne\seqn@m \make@eq{\char\seqn@m}}
\def\SUBEQ{\num@lett\@SUBEQ}
\def\@SUBEQ#1{\subeq \def@name\e@names#1{\char\the\seqn@m}}

\def\SUBEQNBEGIN{\eqtag\SUBEQBEGIN}

\def\eqapp{\num@lett\@eqapp}
\def\@eqapp#1#2{(\fulltag#1#2)}

\def\queq{\num@lett\@queq}
\def\@queq#1{\quote@all{\eq@@fs(#1)}}
\def\qeq{\case@abbr\eqabbr\queq}
\def\qeqs{\case@abbr\eqsabbr\queq}

\outer\def\eqrestore{\all@restore\e@names}

\message{storage management,}



\newtoks\toks@store
\newtoks\file@list \file@list={1234567}
\def\@tmp{\jobname.$$}
\let\ext@ft=F

\begingroup
  \let\storebox=\relax \let\refnam=\relax  
  \let\RFfile=\relax \let\RFext=\relax     
  \newhelp\opt@help{The options \string\refnam, \string\RFfile\space
    and \string\RFext\space are incompatible with \string\storebox.
    Your request will be ignored.}
  \global\opt@help=\opt@help 
  \gdef\opt@err{{\errhelp\opt@help \errmessage{Incompatible options}}}
\endgroup

\begingroup
  \let\storebox=\relax \let\storelist=\relax  
  \let\storefile=\relax \let\RFfile=\relax    
  \gdef\case@store{%
    \glet\storebox\undefined
    \if B\store@blf \glet\storebox\empty \glet\case@store\case@box
      \else \glet\box@store\undefined
      \glet\box@out\undefined \glet\box@print\undefined
      \glet\box@save\undefined \glet\box@kill\undefined \fi
    \glet\storelist\undefined
    \if L\store@blf \glet\storelist\empty \glet\case@store\case@list
      \else \glet\list@store\undefined
      \glet\list@out\undefined \glet\list@print\undefined
      \glet\list@save\undefined \glet\list@kill\undefined \fi
    \glet\storefile\undefined
    \if F\store@blf \glet\storefile\empty \glet\case@store\case@file
      \else \store@setup
      \glet\file@out\undefined \glet\file@print\undefined
      \glet\filef@rm@t\undefined \glet\fil@f@rm@t\undefined
      \glet\file@save\undefined \glet\file@kill\undefined \fi
    \glet\case@box\undefined \glet\case@list\undefined
    \glet\case@file\undefined \glet\store@setup\undefined
    \case@store}
  \gdef\store@setup{\ifx \RFfile\undefined \glet\file@store\undefined
    \glet\file@open\undefined \glet\file@close\undefined
    \glet\file@wlog\undefined \glet\file@free\undefined
    \glet\file@copy\undefined \glet\file@read\undefined \fi}
\endgroup

\outer\def\storebox{\if T\ext@ft \opt@err \else
    \if L\RF@lfe \if N\ref@sbn \opt@err
    \else \glet\store@blf B\fi \else \opt@err \fi \fi}
\outer\def\storelist{\glet\store@blf L}
\outer\def\storefile{\glet\store@blf F}

\def\read@store{\bgroup \@read@store}
\def\read@@store{\bgroup \catcode`\@\l@tter \@read@store}
\def\@read@store#1{\def\after@read{\egroup \toks@store\toks@i
    #1\after@read \ignorespaces}%
  \catcode`\^^M\active \afterassignment\after@read \global\toks@i}
\def\afterread#1{\bgroup \def\after@read{\egroup #1\after@read}}
\let\after@read=\relax
\begingroup
  \catcode`\:=\active
  \gdef\write@save#1{\write@store{:restore\@type{#1}}}
\endgroup
\def\write@store#1{\s@ve{#1{\the\toks@store}}}

\def\f@rm@t#1#2{\bgroup \ignorefoot
  \leftskip\z@skip \rightskip\z@skip \f@rmat
  \ifx @#1@\everypar{\b@format}\else
    \varitem\@indent{#1}\b@format \fi #2\e@format}
\let\f@rmat=\nointerlineskip
\def\form@t{\unskip \strut \par \@break \eform@t}
\def\b@format{\glet\e@format\form@t \strut}
\def\eform@t{\egroup \glet\e@format\eform@t}
\let\e@format=\eform@t

\def\@store{\case@store\box@store\list@store\file@store}
\def\@sstore#1#2#3{\par \noindent \bgroup \captionstyle
  \case@abbr#2#3:\enskip \the\toks@store \par \egroup \@store#1{#3.}}
\def\@add#1{\read@store{\@store#1{}}}
\def\@out#1#2#3#4#5{\case@store\box@out\list@out\file@out#1\begingroup
  \if T#2\let\chap@break\superendpage \fi \chap@break
  \chap@init{\case@language#3}%
  \if C\page@ac \skippagenum T\fi
  \page@reset6#1%
  \@style \he@d{\strut\case@language#4}\@break \@print#1%
  \ifx\chap@break\superendpage \superendpage \fi
  \def\\##1{\glet##1\undefined}\the#5\global#5\emptyt@ks
  \endgroup \fi}
\def\@print{\case@store\box@print\list@print\file@print}
\def\@save{\case@store\box@save\list@save\file@save}
\def\@kill{\case@store\box@kill\list@kill\file@kill}
\def\@ext#1#2#3 {\if B\store@blf \opt@err \else
  \glet\ext@ft T\@add#1{#2#3 }\bgroup
   \def\@store##1##2{}\input#3 \egroup \fi}
\def\@@ext#1#2{\let#1#2\everypar\emptyt@ks
  \def\read@store##1{\relax##1}\def\@store##1{\f@rm@t}\input}

\def\case@box#1#2#3{\case@@store#1%
  \fig@box\tab@box\ref@box\toc@box\foot@box}
\def\case@list#1#2#3{\case@@store#2%
  \fig@list\tab@list\ref@list\toc@list\foot@list}
\def\case@file#1#2#3{\case@@store{\expandafter#3}%
  \fig@file\tab@file\ref@file\toc@file\foot@file}

\def\case@@store#1#2#3#4#5#6#7{\ifcase#7%
  \toks@{\fig@type#1#2}\or
  \toks@{\tab@type#1#3}\or
  \toks@{\ref@type#1#4}\or
  \toks@{\toc@type#1#5}\or
  \toks@{\foot@type#1#6}\fi
  \expandafter\let\expandafter\@type\the\toks@}
\def\@style{\csname\@type style\endcsname}
\def\@indent{\csname\@type indent\endcsname}
\def\@break{\csname\@type break\endcsname}

\def\box@store#1#2{\global\setbox#1\vbox
  {\ifvbox#1\unvbox#1\fi \@style \f@rm@t{#2}{\the\toks@store}}}
\def\box@out{\ifvbox}
\def\box@print{\vskip\baselineskip \unvbox}
\begingroup
  \catcode`\:=\active \catcode`\;=\active
  \gdef\box@save#1{\wlog{; Unable to save text for
    \@type's with option :storebox}}
\endgroup
\def\box@kill#1{{\setbox\z@\box#1}}

\def\list@store#1#2{\toks@\expandafter{#1\\}%
  \xdef#1{\the\toks@ {#2}{\the\toks@store}}}
\def\list@out#1{\ifx #1\empty \else}
\def\list@print#1{\let\\\f@rm@t #1\glet#1\empty}
\def\list@save{\def\\##1##2{\toks@store{##2}\write@save{##1}}%
  \newlinechar`\^^M}
\def\list@kill#1{\glet#1\empty}

\begingroup
  \catcode`\:=\active
  \gdef\file@store#1#2#3#4{%
    \if0#3\expandafter\file@open\the\file@list\@@#1#2\fi
    {\newlinechar`\^^M\let\save@write#2\write@store{::{#4}}}}
\endgroup
\def\file@open#1#2\@@#3#4{\immediate\openout#4\@tmp#1
  \gdef#3{#3#4#1}\global\file@list{#2}\file@wlog{open}#1}
\def\file@wlog#1#2{\wlog{#1 \@tmp#2 for \@type's}}
\def\file@out#1#2#3{\if0#3\else}
\def\file@print#1#2#3{\file@close#2#3\let\\\filef@rm@t
  \file@copy#1#2#3}
\def\filef@rm@t#1{\bgroup \catcode`\@\l@tter \fil@f@rm@t{#1}}
\def\fil@f@rm@t#1#2{\egroup \f@rm@t{#1}{#2}}
\def\file@save#1#2#3{\if0#3\else \file@close#2#3%
  \def\\##1{\read@@store{\expandafter\file@store#1{##1}%
    \write@save{##1}}}%
  \newlinechar`\^^M\file@copy#1#2#3\fi}
\def\file@kill#1#2#3{\if0#3\else \file@close#2#3\file@free#1#2#3\fi}
\def\file@close#1#2{\immediate\closeout#1\file@wlog{close}#2}
\def\file@copy#1#2#3{\file@free#1#2#3\file@read#3}
\def\file@read#1{\input\@tmp#1 }
\def\file@free#1#2#3{\gdef#1{#1#20}%
  \global\file@list\expandafter{\the\file@list#3}}

\message{figures,}


\def\if@t#1#2#3#4#5#6{\glet#1#6\gdef#2{\dot@pref}\glet#3#5\glet#4\relax
  \if#6A\glet#2\empty \glet#3\relax \fi
  \if#6S\gdef#2{\sect@pref}\glet#4#5\fi}
\def\bf@t#1{\let\f@t#1\num@l@tt}
\def\ef@t#1#2#3#4#{\ifx @#4@\g@ne#1\xdef\thef@tn@m{#2\the#1}\else
  \gdef\thef@tn@m{#4}\fi #3\read@store\f@t}
\def\@extf@t#1#2#{\gdef\thef@tn@m{#1}\f@t}


\newtoks\f@names   \f@names={}
\newcount\fign@m   \fign@m=0
\def\fig@type{fig}
\newbox\fig@box
\let\fig@list=\empty
\newwrite\fig@write  \def\fig@file{\fig@file\fig@write0}

\outer\def\figall{\fig@init A}
\outer\def\figchap{\fig@init C}
\outer\def\figsect{\fig@init S}
\def\fig@init{\if@t\fig@acs\fig@pref\fig@reset\fig@@reset\figz@}
\def\figz@{\gz@\fign@m}

\outer\def\figpage{\glet\fig@page T}
\outer\def\nofigpage{\glet\fig@page F}

\def\fig{\bf@t\fig@\@fig}
\def\FIG{\bf@t\fig@\@FIG}
\def\ffig{\bf@t\ffig@\@fig}
\def\FFIG{\bf@t\ffig@\@FIG}

\def\@fig{\ef@t\fign@m\fig@pref\relax}
\def\@FIG#1{\ef@t\fign@m\fig@pref{\def@name\f@names#1{\thef@tn@m}}}

\def\fig@{\@store0{\thef@tn@m .}}
\def\ffig@{\@sstore0\figpref{\thef@tn@m}}
\def\figadd{\@add0}

\def\qufig{\case@abbr\figabbr\num@lett\quote@all}
\def\qufigs{\case@abbr\figsabbr\num@lett\quote@all}

\outer\def\figout{\@out0\fig@page\figpref\fighead\emptyt@ks}
\outer\def\figkill{\@kill0}
\outer\def\restorefig#1{\read@store{\@store0{#1}}}
\outer\def\figrestore{\all@restore\f@names}

\outer\def\FIGext{\@ext0\FIG@ext}
\def\FIG@ext{\@@ext\@FIG\@extf@t}


\newdimen\spictskip  \spictskip=2.5pt

\def\pict{\bf@t\pict@\@fig}
\def\PICT{\bf@t\pict@\@FIG}

\def\pict@{\vskip\the\toks@store \bpargroup
  \raggedright \captionstyle \varitem{\qufig{\thef@tn@m\,}:}%
  \let\@spict\spict@ \spacefactor998\ignorespaces}

\def\spict#1{\ifnum\spacefactor=998\else \parvskip\spictskip \fi
  \@spict{#1\enskip}\ignorespaces}
\def\spict@#1{\setbox\z@\hbox{#1}\advance\hangindent\wd\z@
  \box\z@ \let\@spict\llap}


\outer\def\graphics{\glet\graphic\gr@phic}
\outer\def\nographics{\glet\graphic\nogr@phic}

\def\gr@phic#1{\vbox\bgroup \def\gr@@@ph{#1}\bfr@me\gr@ph}
\def\nogr@phic#1{\vbox\bgroup \write\m@ne{Insert plot #1}\bfr@me\fr@me}
\def\frame{\vbox\bgroup \bfr@me\fr@me}
\def\bfr@me#1#2#3#4{\tfr@me\z@\dimen@iv#2\relax 
  \dimen@#3\relax \tfr@me\dimen@\dimen@ii#4\relax 
  \setbox\z@\hbox to\dimen@iv{#1}\ht\z@\dimen@ \dp\z@\dimen@ii
  \box\z@ \egroup}
\def\tfr@me#1#2#3\relax{#2#3\relax \ifdim#2<-#1\errhelp\fr@mehelp
  \errmessage{Invalid box size}#2-#1\fi}
\newhelp\fr@mehelp{The \string\wd\space and \string\ht+\string\dp\space
  of a \string\frame\space or \string\graphic\space \string\box\space
  must not be negative and will be changed to 0pt.}

\def\gr@ph{\lower\dimen@ii\gr@@ph b\hfil
  \raise\dimen@\gr@@ph e}
\def\gr@@ph#1{\hbox{\special{^X\gr@@@ph^A}}}         
\def\gr@@ph#1{\hbox{\special{#1plot GKSM \gr@@@ph}}} 
\def\gr@@ph#1{\hbox{\special{^X#1plot GKSM \gr@@@ph^A}}} 
\def\fr@me{\vrule\fr@@@me
  \bgroup \dimen@ii-\dimen@ \fr@@me\dimen@ii \hfilneg
  \bgroup \dimen@-\dimen@ii \fr@@me\dimen@ \vrule\fr@@@me}
\def\fr@@me#1{\advance#1.4\p@ \leaders \hrule\fr@@@me \hss \egroup}
\def\fr@@@me{\@height\dimen@ \@depth\dimen@ii}

\message{tables,}


\newtoks\t@names   \t@names={}
\newcount\tabn@m   \tabn@m=0
\def\tab@type{tab}
\newbox\tab@box
\let\tab@list=\empty
\newwrite\tab@write  \def\tab@file{\tab@file\tab@write0}

\outer\def\taball{\tab@init A}
\outer\def\tabchap{\tab@init C}
\outer\def\tabsect{\tab@init S}
\def\tab@init{\if@t\tab@acs\tab@pref\tab@reset\tab@@reset\tabz@}
\def\tabz@{\gz@\tabn@m}

\outer\def\tabpage{\glet\tab@page T}
\outer\def\notabpage{\glet\tab@page F}

\def\tab{\bf@t\tab@\@tab}
\def\TAB{\bf@t\tab@\@TAB}
\def\ttab{\bf@t\ttab@\@tab}
\def\TTAB{\bf@t\ttab@\@TAB}

\def\@tab{\ef@t\tabn@m\tab@pref\relax}
\def\@TAB#1{\ef@t\tabn@m\tab@pref{\def@name\t@names#1{\thef@tn@m}}}

\def\tab@{\@store1{\thef@tn@m .}}
\def\ttab@{\@sstore1\tabpref{\thef@tn@m}}
\def\tabadd{\@add1}

\def\qutab{\case@abbr\tababbr\num@lett\quote@all}
\def\qutabs{\case@abbr\tabsabbr\num@lett\quote@all}

\outer\def\tabout{\@out1\tab@page\tabpref\tabhead\emptyt@ks}
\outer\def\tabkill{\@kill1}
\outer\def\restoretab#1{\read@store{\@store1{#1}}}
\outer\def\tabrestore{\all@restore\t@names}

\outer\def\TABext{\@ext1\TAB@ext}
\def\TAB@ext{\@@ext\@TAB\@extf@t}


\newskip\htabskip   \htabskip=1em plus 2em minus .5em
\newdimen\vtabskip  \vtabskip=2.5pt
\newbox\tab@top   \newbox\tab@bot

\let\@hrule=\hrule
\let\@halign=\halign
\let\@valign=\valign
\let\@span=\span
\let\@omit=\omit

\def\@@span{\@span\@omit\@span}
\def\@@@span{\@span\@omit\@@span}

\def\sp@n{\span\@omit\advance\mscount\m@ne} 

\def\table#1#{\vbox\bgroup\offinterlineskip
  \toks@ii{#1\bgroup \unhcopy\tab@top \unhcopy\tab@bot
    ##}\afterassignment\tab@preamble \@eat}


\def\tab@preamble#1\cr{\let\tab@@vrule\tab@repeat
  \let\tab@amp@\empty \let\tab@amp\empty \let\span@\@@span
  \toks@{\tab@space#1&\cr}\the\toks@}
\def\tab@space{\tab@test{ }{}\tab@vrule}                 
\def\tab@vrule{\tab@test\vrule{\tab@add\vrule}
  \tab@@vrule}
\def\tab@repeat{\tab@test&{\tab@add&
    \let\tab@@vrule\tab@template \let\tab@amp@\tab@@amp
    \let\tab@amp&\let\span@\@@@span}\tab@template}
\def\tab@template#1&{\tab@add{\@span\tab@amp@
    \tabskip\htabskip&\tab@setup#1&\tabskip\z@skip##}
  \tab@test\cr{\let\tab@@vrule\tab@exec}\tab@space}      
\def\tab@@amp{&##}

\def\tab@test#1#2#3{\let\tab@comp= #1\toks@{#2}\let\tab@go#3%
  \futurelet\n@xt \tab@@test}
\def\tab@@test{\ifx \tab@comp\n@xt \the\toks@
    \afterassignment\tab@go \expandafter\@eat \else
  \expandafter\tab@go \fi}

\def\tab@add#1{\toks@ii\expandafter{\the\toks@ii#1}}


\def\tab@exec{\tab@r@set\everycr{\tab@body}%
  \def\halign{\tab@r@set \halign}\def\valign{\tab@r@set \valign}%
  \def\omit{\@omit \tab@setup}%
  \def\n@xt##1{\hbox{\dimen@\ht\strutbox\dimen@ii\dp\strutbox
    \advance##1\vtabskip \vrule \@height\dimen@ \@depth\dimen@ii
    \@width\z@}}%
  \setbox\tab@top\n@xt\dimen@ \setbox\tab@bot\n@xt\dimen@ii
  \def\ml##1{\relax                                      
    \ifmmode \let\@ml\empty \else \let\@ml$\fi           
    \@ml\vcenter{\hbox\bgroup\unhcopy\tab@top            
    ##1\unhcopy\tab@bot\egroup}\@ml}%
  \def\nl{\egroup\hbox\bgroup\strut}
  \tabskip\z@skip\@halign\the\toks@ii\cr}

\def\tab@r@set{\let\cr\endline \everycr\emptyt@ks
  \let\halign\@halign \let\valign\@valign
  \let\span\@span \let\omit\@omit}
\def\tab@setup{\relax \iffalse {\fi \let\span\span@ \iffalse }\fi}


\def\tab@body{\noalign\bgroup \tab@@body}           
\def\tab@@body{\futurelet\n@xt \tab@end}            
\def\tab@end{\ifcat\egroup\noexpand\n@xt
    \expandafter\egroup \expandafter\egroup \else        
  \expandafter\tab@blank \fi}
\def\tab@blank{\ifcat\space\noexpand\n@xt
    \afterassignment\tab@@body \expandafter\@eat \else   
  \expandafter\tab@hrule \fi}
\def\tab@hrule{\ifx\hrule\n@xt
    \def\hrule{\@hrule\egroup \tab@body}\else            
  \expandafter\tab@noalign \fi}
\def\tab@noalign{\ifx\noalign\n@xt
    \aftergroup\tab@body \expandafter\@eat@ \else        
  \expandafter\tab@row \fi}
\def\tab@row#1\cr{\toks@ii{\toks@ii{\egroup}
    \tab@item#1&\cr}\the\toks@ii}
\def\tab@item#1&{\tab@add{\tab@amp&#1&}
  \futurelet\n@xt \tab@cr}
\def\tab@cr{\ifx\cr\n@xt
    \expandafter\the\expandafter\toks@ii \else           
  \expandafter\tab@item \fi}

\message{references,}


\newtoks\r@names   \r@names={}
\newcount\refn@m   \refn@m=0
\newcount\ref@temp
\def\ref@type{ref}
\newbox\ref@box
\let\ref@list=\empty
\newwrite\ref@write  \def\ref@file{\ref@file\ref@write0}

\begingroup \let\refnam=\relax  
  \newhelp\ref@help{The option \string\refnam\space allows predefined
    references only and is incompatible with \string\qref(s).
    Your request will be ignored.}
  \global\ref@help=\ref@help 
  \gdef\ref@err{{\errhelp\ref@help \errmessage{Invalid request}}}
\endgroup

\begingroup
  \let\refsup=\relax \let\refsqb=\relax  
  \let\refnam=\relax                     
  \gdef\ref@setup{%
    \glet\refsup\undefined
    \if S\ref@sbn \glet\refsup\empty \glet\therefn@m\suprefn@m \fi
    \glet\refsqb\undefined
    \if B\ref@sbn \glet\refsqb\empty \glet\therefn@m\sqbrefn@m \fi
    \glet\refnam\undefined
    \if N\ref@sbn \glet\refnam\empty \glet\the@quref\nam@quref
      \glet\@@ref\ref@err \gdef\qref{\ref@err \quref}\glet\qrefs\qref
      \glet\RF@def@\RF@def@nam
      \glet\RF@find\undefined \glet\RF@search\undefined
      \glet\RF@locate\undefined \glet\@RFread\undefined
      \glet\qurefsup\ref@err \glet\sup@quref\undefined
      \glet\qurefsqb\ref@err \glet\sqb@quref\undefined
      \glet\qurefnum\ref@err \glet\num@quref\undefined
      \glet\ref@restore\ref@err
      \else \gdef\@@ref{\@store2\therefn@m}%
      \gdef\qref{\case@abbr\refabbr\num@lett\quote@all}%
      \gdef\qrefs{\case@abbr\refsabbr\num@lett\quote@all}%
      \glet\RF@def@\RF@def@num \glet\RF@print\undefined
      \gdef\ref@restore{\all@restore\r@names}\fi
    \glet\nam@quref\undefined
    \gdef\quref{\ref@unskip \num@lett\the@quref}%
    \glet\RF@def@num\undefined \glet\RF@def@nam\undefined
    \gdef\RF@restore{\all@restore\R@names}%
    \glet\ref@setup\undefined}
\endgroup

\outer\def\refsup{\glet\ref@sbn S\global\qurefsup}
\outer\def\refsqb{\glet\ref@sbn B\global\qurefsqb}
\outer\def\refnam{\if B\store@blf \opt@err \else \glet\ref@sbn N\fi}

\def\qurefsup{\let\the@quref\sup@quref}
\def\qurefsqb{\let\the@quref\sqb@quref}
\def\qurefnum{\let\the@quref\num@quref}

\outer\def\refpage{\glet\ref@page T}
\outer\def\norefpage{\glet\ref@page F}

\outer\def\refkeep{\glet\ref@kc K}
\outer\def\refclear{\glet\ref@kc C}

\def\ref{\ref@advance \refend \@ref}
\def\REF{\num@lett\@REF}
\def\@REF#1{\ref@name#1\@ref}
\def\refend{\quref{\the\refn@m}}

\def\refs{\ref@advance \ref@temp\refn@m \@ref}
\def\REFS{\num@lett\@REFS}
\def\@REFS#1{\ref@name#1\ref@temp\refn@m \@ref}
\def\refscon{\ref@advance \@ref}

\def\refsend{\quref{\the\ref@temp -\the\refn@m}}

\def\ref@advance{\ref@unskip \g@ne\refn@m}
\def\ref@name{\ref@@name\r@names}
\def\ref@@name#1#2{\ref@advance \def@name#1#2{\the\refn@m}}
\def\ref@unskip{\ifhmode \unskip \fi}

\def\suprefn@m{\the\refn@m .}
\def\sqbrefn@m{$\lbrack \the\refn@m \rbrack$}
\def\@ref{\read@store\@@ref}
\def\@@ref{\ref@setup \@@ref}
\def\refadd{\@add2}

\def\sup@quref#1{\leavevmode \nobreak \quote@all{^{#1}}}
\def\sqb@quref#1{\ \quote@all{\lbrack #1\rbrack}}
\def\num@quref{\ \quote@all}
\def\nam@quref{\@use}
\def\quref{\ref@setup \quref}
\def\qref{\ref@setup \qref}
\def\qrefs{\ref@setup \qrefs}

\outer\def\refout{{\if K\ref@kc \@out2\ref@page\refpref\refhead\emptyt@ks
  \else \@out2\ref@page\refpref\refhead\r@names
  \let\\\@RFdef \the\R@names \global\R@names\emptyt@ks
  \if L\RF@lfe \else \glet\RF@list\empty \gz@\RF@high \fi
  \gz@\refn@m \fi}}
\outer\def\refkill{\@kill2}
\outer\def\restoreref#1{\read@@store{\@store2{#1}}}
\outer\def\refrestore{\ref@restore}
\outer\def\RFrestore{\RF@restore}
\def\ref@restore{\ref@setup \ref@restore}
\def\RF@restore{\ref@setup \RF@restore}

\outer\def\REFext{\@ext2\REF@ext}
\def\REF@ext{\@@ext\ref@name\refn@m}


\newtoks\R@names   \R@names={}
\newcount\RFn@m   \newcount\RF@high
\newcount\RFmax   \RFmax=50  
\def\RF@type{RF}
\let\RF@list=\empty
\newwrite\RF@write  \def\RF@file{\RF@file\RF@write0}
\let\RF@noc=N

\begingroup \let\storefile=\relax        
  \let\RFlist=\relax \let\RFfile=\relax  
  \let\RFext=\relax \let\RF=\relax       
  \gdef\@RF{%
    \glet\RFlist\undefined
    \if L\RF@lfe \glet\RFlist\empty \glet\@RF\@RFlist
      \gdef\RF@input{\RF@list}%
      \else \glet\@RFlist\undefined \fi
    \glet\RFfile\undefined
    \if F\RF@lfe \glet\RFfile\empty \glet\@RF\@RFfile
      \else \RF@setup
      \glet\@RFfile\undefined \glet\@RFcopy\undefined
      \glet\RF@store\undefined \glet\RF@copy\undefined
      \glet\RF@@input\undefined \fi
    \glet\RFext\undefined
    \if E\RF@lfe \gdef\RFext##1 {}\glet\@RF\@RFext
      \else \glet\@RFext\undefined \fi
    \glet\RF@setup\undefined \@RF}
  \gdef\RF@setup{\ifx \storefile\undefined \glet\file@store\undefined
    \glet\file@open\undefined \glet\file@close\undefined
    \glet\file@wlog\undefined \glet\file@free\undefined
    \glet\file@copy\undefined \glet\file@read\undefined \fi}
\endgroup

\outer\def\RFlist{\glet\RF@lfe L}
\outer\def\RFfile{\if B\store@blf \opt@err \else \glet\RF@lfe F\fi}
\outer\def\RFext#1 {\if B\store@blf \opt@err \else
  \glet\RF@lfe E\gdef\RF@input{\input#1 }\RF@input \fi}

\def\@RFdef#1{\gdef#1{\RF@def#1}}
\def\@RF@list#1{\toks@\expandafter{\RF@list\RF@#1}%
  \xdef\RF@list{\the\toks@ {\the\toks@store}}}
\def\@RFcopy#1{\RF@store{\noexpand#1}}
\def\RF@store{\let\@type\RF@type \if C\RF@noc \glet\RF@noc N%
  \RF@copy \fi \glet\RF@noc O\expandafter\file@store\RF@file}
\def\RF@input{\expandafter\RF@@input\RF@file}
\begingroup \let\RF=\relax  
  \gdef\RF@copy{{\let\\\RF \let\@RF\@RFcopy
    \expandafter\file@copy\RF@file}}
  \gdef\RF@@input#1#2#3{\if O\RF@noc \let\@type\RF@type
    \file@close#2#3\glet\RF@noc C\fi \let\\\RF \file@read#3}
\endgroup

\def\RF@def#1{\ref@@name\R@names#1\RF@def@ #1}
\def\RF@def@{\ref@setup \RF@def@}
\def\RF@def@num{\toks@store\expandafter{\expandafter\RF@find
  \expandafter{\the\refn@m}}\@@ref}
\def\RF@def@nam{\ifnum\refn@m=\@ne \refadd{\RF@print}\fi}
\def\RF@test#1{\z@}

\def\RF@print{\let\@RF\@RF@print \let\RF@def\RF@test
  \let\RF@first T\RF@input}
\def\@RF@print#1{\ifnum#1>\z@
  \if\RF@first T\let\RF@first F\setbox\z@\lastbox
  \else \form@t\f@rmat \noindent \strut \fi
  \hangindent\namrefindent \the\toks@store \fi}

\def\RF@find#1{\RFn@m#1\bgroup \let\RF@def\RF@test
  \if L\RF@lfe \else \ifnum\RFn@m<\RF@high \else \RF@search \fi \fi
  \let\RF@\RF@locate \RF@list \egroup}

\def\RF@search{\global\RF@high\RFn@m \global\advance\RF@high\RFmax
  \glet\RF@list\empty \let\@RF\@RFread \let\par\relax \RF@input}
\def\@RFread#1{\ifnum#1<\RFn@m \else \ifnum#1<\RF@high
  \@RF@list#1\fi \fi}
\def\RF@locate#1#2{\ifnum#1=\RFn@m #2\fi}

\def\@RFlist#1{\@RFdef#1\@RF@list#1}
\def\@RFfile#1{\@RFdef#1\@RFcopy#1}
\let\@RFext=\@RFdef

\outer\def\RF{\num@lett\RF@}
\def\RF@#1{\ref@unskip \read@store{\@RF#1}}


\outer\def\yearpage{\glet\yearpage@yp Y}
\outer\def\pageyear{\glet\yearpage@yp P}

\def\journal#1{{\journalstyle{#1}}\j@urnal{}}
\def\journalp#1{{\journalstyle{#1}}\j@urnal}
\def\journalf#1#2#3({{\journalstyle{#1}}\j@urnal{#3}#2(}

\def\j@urnal#1#2(#3)#4*{\unskip
  \ {\volumestyle{#1\ifx @#1@\else\ifx @#2@\else
  \kern.2em\fi \fi#2}}\unskip
  \ifx @#3@\else\ifx @#4@ (#3)\else\if Y\yearpage@yp\ (#3) #4\else
  , #4 (#3)\fi \fi \fi}

\def\Phys{Phys.\ }
\def\Rev{Rev.\ }

\def\PRD{\journalp{\Phys\Rev}D}


\message{table of contents,}


\newcount\tocn@m   \tocn@m=0
\newcount\auto@toc   \auto@toc=-1
\let\toc@saved\empty

\def\toc@type{toc}
\newbox\toc@box
\let\toc@list=\empty
\newwrite\toc@write  \def\toc@file{\toc@file\toc@write0}

\outer\def\tocpage{\glet\toc@page T}
\outer\def\notocpage{\glet\toc@page F}

\outer\def\tocnone{\gm@ne\auto@toc}
\outer\def\tocchap{\gz@\auto@toc}
\outer\def\tocsect{\global\auto@toc\@ne}

\def\toc#1{\read@store{\@toc{#1}}}
\def\tocadd{\@add3}
\def\@toc{\g@ne\tocn@m
  \expandafter\@@toc\csname toc@\romannumeral\tocn@m\endcsname}
\def\@@toc#1{\pagelabel#1%
  \toks@store\expandafter{\the\toks@store\toc@fill#1}\@store3}
\def\toc@fill{\rightskip4em\@plus1em\@minus1em\parfillskip-\rightskip
  \unskip\vadjust{}\leaders\hbox to1em{\hss.\hss}\hfil}

\outer\def\tocout{\@out3\toc@page\tocpref\tochead\emptyt@ks}
\outer\def\tockill{\@kill3}
\outer\def\restoretoc#1{\read@@store{\@store3{#1}}}

\message{footnotes,}


\newcount\footn@m   \footn@m=0
\def\foot@type{foot}
\newbox\foot@box
\let\foot@list=\empty
\newwrite\foot@write  \def\foot@file{\foot@file\foot@write0}

\outer\def\footsqb{\glet\foot@bp B\glet\thefootn@m\sqbfootn@m}
\outer\def\footpar{\glet\foot@bp P\glet\thefootn@m\parfootn@m}

\outer\def\footbot{\glet\foot@be B\glet\vfootnote\vfootn@te}
\outer\def\footend{\glet\foot@be E\glet\vfootnote\foot@store}

\outer\def\footpage{\glet\foot@page T}
\outer\def\nofootpage{\glet\foot@page F}

\def\sqbfootn@m{\lbrack \the\footn@m \rbrack}
\def\parfootn@m{\the\footn@m )}

\def\foot{\hfoot \vfootnote\footid}
\def\hfoot{\g@ne\footn@m \edef\n@xt{{$^{\thefootn@m}$}}%
  \expandafter\hfootnote\n@xt}
\def\footnote#1{\hfootnote{#1}\vfootnote\footid}
\def\hfootnote#1{\let\@sf\empty
  \ifhmode\unskip\edef\@sf{\spacefactor\the\spacefactor}\/\fi
  #1\@sf \gdef\footid{#1}}
\def\vfootn@te#1{\insert\footins\bgroup \foot@style
  \llap{#1}\after@arg\@foot}
\let\fo@t=\undefined
\let\f@@t=\undefined
\let\f@t=\undefined

\def\foot@style{\footstyle  
  \interlinepenalty\interfootnotelinepenalty
  \baselineskip\footnotebaselineskip
  \splittopskip\interfootnoteskip 
  \splitmaxdepth\dp\strutbox \floatingpenalty\@MM
  \leftskip.05\hsize \rightskip\z@skip
  \spaceskip\z@skip \xspaceskip\z@skip \noindent \footstrut}

\def\foot@store#1{\read@store{\@store4{#1}}}
\def\footadd{\@add4}

\outer\def\footout{\@out4\foot@page\footpref\foothead\emptyt@ks}
\outer\def\footkill{\@kill4}
\outer\def\restorefoot#1{\read@store{\@store4{#1}}}

\def\ignorefoot{\let\foot\eat \let\hfoot\eat  
  \def\footnote{\expandafter\eat\eat}\let\hfootnote\footnote}

\message{items and points,}



\def\varitem{\afterassignment\v@ritem \setbox\z@\hbox}
\def\v@ritem{\hss \bgroup \aftergroup\v@@ritem}
\def\v@@ritem{\enskip \egroup \endgraf \noindent
  \hangindent\wd\z@ \box\z@ \ignorespaces}

\def\hvskip{\afterassignment\h@vskip \skip@}
\def\h@vskip{\unskip\nobreak \vadjust{\vskip\skip@}\lb \ignorespaces}

\def\parvskip{\bgroup \afterassignment\par@vskip \parskip}
\def\par@vskip{\parindent\hangindent \endgraf \indent \egroup
  \ignorespaces}

\def\item{\varitem to2.5em}
\def\sitem{\varitem to4.5em}
\def\ssitem{\varitem to6.5em}


\newcount\pointn@m   \pointn@m=0
\def\pointbegin{\gz@\pointn@m \point}
\def\point{\g@ne\pointn@m
  \xdef\the@label{\the\pointn@m}\item{\the@label.}}

\newcount\spointn@m   \spointn@m=96
\def\spointbegin{\global\spointn@m96 \spoint}
\def\spoint{\g@ne\spointn@m
  \xdef\the@label{\char\the\spointn@m}\sitem{(\the@label)}}

\newcount\sspointn@m   \sspointn@m=0
\def\sspointbegin{\gz@\sspointn@m \sspoint}
\def\sspoint{\g@ne\sspointn@m
  \xdef\the@label{\romannumeral\sspointn@m}\ssitem{\the@label)}}



\message{matrices and additional math symbols,}


\def\matc{\let\mat@lfil\hfil \let\mat@rfil\hfil}
\def\matl{\let\mat@lfil\relax \let\mat@rfil\hfil}
\def\matr{\let\mat@lfil\hfil \let\mat@rfil\relax}

\def\matrix#1{\null\,\vcenter{\normalbaselines\m@th
    \ialign{$\mat@lfil##\mat@rfil$&&\quad$\mat@lfil##\mat@rfil$\crcr
      \mathstrut\crcr\noalign{\kern-\baselineskip}%
      #1\crcr\mathstrut\crcr\noalign{\kern-\baselineskip}}}\,}

\def\bordermatrix#1{\begingroup \m@th
  \setbox\z@\vbox{%
    \def\cr{\crcr\noalign{\kern2\p@\glet\cr\endline}}%
    \ialign{$##\hfil$\kern2\p@\kern\p@renwd&\thinspace$\mat@lfil##%
      \mat@rfil$&&\quad$\mat@lfil##\mat@rfil$\crcr
      \omit\strut\hfil\crcr\noalign{\kern-\baselineskip}%
      #1\crcr\omit\strut\cr}}%
  \setbox\tw@\vbox{\unvcopy\z@\global\setbox\@ne\lastbox}%
  \setbox\tw@\hbox{\unhbox\@ne\unskip\global\setbox\@ne\lastbox}%
  \setbox\tw@\hbox{$\kern\wd\@ne\kern-\p@renwd\left(\kern-\wd\@ne
    \global\setbox\@ne\vbox{\box\@ne\kern2\p@}%
    \vcenter{\kern-\ht\@ne\unvbox\z@\kern-\baselineskip}\,\right)$}%
  \null\;\vbox{\kern\ht\@ne\box\tw@}\endgroup}


\mathchardef\smallsum=\dq1006
\mathchardef\smallprod=\dq1005

\def\b@mmode{\relax\ifmmode \expandafter\c@mmode \else $\fi}
\def\c@mmode#1\e@mmode{#1}
\def\e@mmode{$}
\def\defmmode#1#2{\def#1{\b@mmode#2\e@mmode}}

\defmmode\{{\lbrace}
\defmmode\}{\rbrace}

\defmmode\,{\mskip\thinmuskip}
\defmmode\>{\mskip\medmuskip}
\defmmode\;{\mskip\thickmuskip}

\defmmode{\Mit#1}{\mit#1}
\defmmode{\Cal#1}{\cal\uppercase\expandafter{#1}}

\def\dotii#1{{\mathop{#1}\limits^{\vbox to -1.4\p@{\kern-2\p@
   \hbox{\tenrm..}\vss}}}}
\def\dotiii#1{{\mathop{#1}\limits^{\vbox to -1.4\p@{\kern-2\p@
   \hbox{\tenrm...}\vss}}}}
\def\dotiv#1{{\mathop{#1}\limits^{\vbox to -1.4\p@{\kern-2\p@
   \hbox{\tenrm....}\vss}}}}

\let\barsymbol -
\mathchardef\tildesymbol=\dq0218
\def\hatsymbol{{\mathchoice{\null}{\null}{\,\,\hbox{\lower 10\p@\hbox
    {$\widehat{\null}$}}}{\,\hbox{\lower 20\p@\hbox
       {$\hat{\null}$}}}}}


\def\begin@stmt{\par\noindent\bpargroup\stmttitlestyle}
\def\adv@stmt#1#2#3#4{\begin@stmt
  \count@\ifx#2#3#1 \else\z@ \glet#2#3\fi \advance\count@\@ne
  \xdef#1{\the\count@}\edef\the@label{#4#1}}

\def\make@stmt{\ \the@label \make@@stmt}
{\catcode`\:\active
  \gdef\make@@stmt{\ \catcode`\:\active \let:\end@stmt}
}
\def\end@stmt{\catcode`\:\@ther \unskip :\stmtstyle
  \enskip \ignorespaces}

\def\defstmt#1#2#3{\expandafter\def@stmt \csname#1\endcsname
  {#2}{#1@stmt@}#3@}
\def\def@stmt#1#2#3#4#5@{\bgroup
  \toks@{\begin@stmt #2\make@@stmt}\toks@ii{#2\make@stmt}%
  \if#4n\xdef#1{\the\toks@}\else
    \xdef#1{\csname#3adv\endcsname \the\toks@ii}%
    \if#4=\edef\n@xt{\expandafter\noexpand
      \csname#5@stmt@adv\endcsname}\else
      \toks@{\empty}\toks@ii\toks@ \if#4c\toks@{\dot@pref}\fi
      \if#4s\toks@{\sect@pref}\if#5c\toks@ii{\sect@dot@pref}\fi \fi
      \if#5a\toks@ii\toks@ \fi
      \edef\n@xt{\noexpand\adv@stmt \csname#3num\endcsname
        \csname#3save\endcsname \the\toks@ \the\toks@ii}\fi
    \expandafter\glet\csname#3adv\endcsname\n@xt \fi
  \egroup}

\def\Prf{\par\noindent\bpargroup\prftitlestyle \case@language\prfhead
  \let\stmtstyle\prfstyle \make@@stmt}

\message{save, restore and start macros,}


\def\save@type{.texsave }  
\newread\test@read
\newwrite\save@write

\def\s@ve{\immediate\write\save@write}
\begingroup \catcode`\:=\active \catcode`\;=\active
  \outer\gdef\save#1 {{\let\,\space
    \immediate\openout\save@write#1\save@type
    \s@ve{;* definitions for :restore #1 \date\space- \thetime\space*}%
    \s@ve{:comment}%
    \s@ve{:mainlanguage:\case@language{german\else english}}%
    \save@page \save@chap
    \save@equ \save@fig \save@tab \save@ref \save@toc \save@foot
    \bgroup \def\n@xt##1.##2{\advance##2\@ne \s@ve{:start##1\the##2}}%
      \n@xt chap.\chapn@m \n@xt sect.\sectn@m
      \n@xt appendix.\appn@m \n@xt equ.\eqn@m
      \n@xt fig.\fign@m \n@xt tab.\tabn@m
      \n@xt ref.\refn@m \n@xt toc.\tocn@m
      \n@xt foot.\footn@m \egroup
    \s@ve{:\ifx\chap@@eq\sect@@eq app\else
      \ifnum\chapn@m=\z@ sect\else chap\fi \fi init}%
    \@save0\@save1\@save2\@save3\@save4%
    \if K\ref@kc \s@ve{:endcomment}\fi
    \save@restore ref \r@names  \save@restore RF \R@names
    \if C\ref@kc \s@ve{:endcomment}\fi
    \save@restore lbl \l@names  \save@restore eq \e@names
    \save@restore fig \f@names  \save@restore tab \t@names
    \s@ve{;*  end of definitions  *}\immediate\closeout\save@write
    }\wlog{* file #1\save@type saved *}}
  \gdef\save@restore#1 {\def\\{\s@ve{:#1restore}%
      \let\\\save@@restore \\}\the}
  \gdef\save@@restore#1{\toks@\expandafter{#1}%
    \s@ve{:dorestore\string#1{\the\toks@}}}
  \gdef\save@page{\s@ve{:\@opt\page@tbn Ttop Bbot Nno *pagenum%
      :\@opt\head@lrac Llef Rrigh Aal Ccen *thead%
      :\@opt\foot@lrac Llef Rrigh Aal Ccen *tfoot%
      :page\@opt\page@ac Aall Cchap *%
      :\@opt\ori@pl Pportrait Llandscape *}%
    \s@ve{:startpage\@opt\page@ac C\the\pageno@pref. *\the\pageno}}
  \gdef\save@chap{\s@ve{:\@opt\chap@page Fno *chappage%
      :\@opt\chap@yn Nno *chapters%
      :\@opt\chap@ar Aarabic Rroman *chapnum}}
  \gdef\save@equ{\s@ve{:equ\@opt\eq@acs Aall Cchap Ssect *%
      :equ\@opt\eq@lrn Lleft Rright Nnone *%
      :equ\@opt\eq@fs Ffull Sshort *}}
  \gdef\save@fig{\s@ve{:\@opt\fig@page Fno *figpage%
      :fig\@opt\fig@acs Aall Cchap Ssect *}}
  \gdef\save@tab{\s@ve{:\@opt\tab@page Fno *tabpage%
      :tab\@opt\tab@acs Aall Cchap Ssect *}}
  \gdef\save@ref{\s@ve{:\@opt\ref@page Fno *refpage%
      :ref\@opt\ref@kc Kkeep Cclear *%
      :ref\@opt\ref@sbn Ssup Bsqb Nnam *%
      :\@opt\yearpage@yp Yyearpage Ppageyear *}}
  \gdef\save@toc{\s@ve{:\@opt\toc@page Fno *tocpage%
      :toc\ifcase\save@@toc \else none\fi}}
  \gdef\save@foot{\s@ve{:\@opt\foot@page Fno *footpage%
      :foot\@opt\foot@be Bbot Eend *%
      :foot\@opt\foot@bp Bsqb Ppar *}}
\endgroup
\def\@opt#1#2#3 #4{\if #1#2#3\fi \if #4*\else
  \expandafter\@opt\expandafter#1\expandafter#4\fi}
\def\save@@toc{\auto@toc chap\or sect}

\newif\ifcr@ss
\begingroup \let\comment=\relax
  \outer\gdef\restore{\@kill0\@kill1\@kill2\@kill3\@kill4%
    {\def\\##1{\glet##1\undefined}%
      \def\n@xt##1{\the##1\global##1\emptyt@ks}%
      \n@xt\l@names \n@xt\e@names \n@xt\f@names \n@xt\t@names
      \n@xt\r@names \let\\\@RFdef \n@xt\R@names}
    \bgroup \let\comment\relax \let\d@rest@re\dorest@re \@restore}
\endgroup
\def\crossrestore#1 {\bgroup \openin\test@read#1\save@type
  \ifeof\test@read \let\n@xt\egroup
    \message{* file #1\save@type missing *}%
  \else \closein\test@read
    \def\n@xt{\@restore#1 }\let\d@rest@re\docr@ss \fi \n@xt}
\def\@restore#1 {\input #1\save@type \egroup}
\def\all@restore{\glet\name@list}
\outer\def\dorestore{\bgroup \catcode`\@\l@tter \num@lett\d@restore}
\def\d@restore#1#2{\egroup \toks@{#2}\d@rest@re#1}
\def\dorest@re#1{\def@name\name@list#1{\the\toks@}}
\def\docr@ss#1{\ifx#1\undefined \cr@sstrue
    \else \expandafter\testcr@ss#1\cr@ss\@@ \fi
  \ifcr@ss \expandafter\testcr@ss\the\toks@\cr@ss\@@ \else \cr@sstrue \fi
  \ifcr@ss\else \expandafter\dorest@re\expandafter#1\fi}
\def\testcr@ss#1\cr@ss#2\@@{\ifx @#2@\toks@{\cr@ss#1}\cr@ssfalse
  \else \cr@sstrue \fi}
\let\cr@ss=\empty

\def\def@name#1#2{\let\name@list#1\add@name#2\xdef#2}
\def\del@name#1{\bgroup
  \def\n@xt##1\\#1##2\\#1##3\@@##4{\global##4{##1##2}}%
  \def\del@@name##1{\expandafter\n@xt\the##1\\#1\\#1\@@##1}%
  \del@@name\l@names \del@@name\e@names \del@@name\f@names
  \del@@name\t@names \del@@name\r@names \del@@name\R@names \egroup}
\def\add@name#1{\del@name#1%
  {\global\name@list\expandafter{\the\name@list\\#1}}}

\def\kill{\num@lett\@k@ll}
\def\@k@ll#1{\def\k@ll##1{\ifx##1\k@ll \let\k@ll\relax \else
    \del@name##1\glet##1\undefined \fi \k@ll}\k@ll#1\k@ll}


\outer\def\startchap{\st@rt\chapn@m}
\outer\def\startsect{\st@rt\sectn@m}
\outer\def\startappendix{\st@rt\appn@m}
\outer\def\startequ{\st@rt\eqn@m}
\outer\def\startfig{\st@rt\fign@m}
\outer\def\starttab{\st@rt\tabn@m}
\outer\def\startref{\st@rt\refn@m}
\outer\def\starttoc{\st@rt\tocn@m}
\outer\def\startfoot{\st@rt\footn@m}

\def\st@rt#1{\gm@ne#1 \global\advance#1}

\message{installation dependent parameters,}




\hoffset@corr@p=86mm   \voffset@corr@p=98mm      
\hoffset@corrm@p=-5mm   \voffset@corrm@p=4mm      
\hoffset@corr@l=134mm   \voffset@corr@l=70mm     
\hoffset@corrm@l=-5mm   \voffset@corrm@l=-4mm     

\catcode`\@=12 

\message{and default options.}



\hbadness=2000  

\newsect        

\mainlanguage\english 

\botpagenum     
\centhead       
\centfoot       
\pageall        
\portrait       
\titlepage      
\nochappage     
\arabicchapnum  
\chapters       
\equchap        
\equshort       
\equright       
\storelist      
\figall         
\figpage        
\nographics     
\taball         
\tabpage        
\refsup         
\refpage        
\refkeep        
\RFlist         
\yearpage       
\tocpage        
\tocnone        
\footsqb        
\footbot        
\footpage       
\matc           

\wlog{summary of allocations:}
\wlog{last count=\number\count10 }
\wlog{last dimen=\number\count11 }
\wlog{last skip=\number\count12 }
\wlog{last muskip=\number\count13 }
\wlog{last box=\number\count14 }
\wlog{last toks=\number\count15 }
\wlog{last read=\number\count16 }
\wlog{last write=\number\count17 }
\wlog{last fam=\number\count18 }
\wlog{last language=\number\count19 }
\wlog{last insert=\number\count20 }

   \def\Fmtversion{2.0}

\edef\fmtversion{\Fmtversion(\fmtname\space\fmtversion)}
\let\fmtname=\Fmtname 
\everyjob={
  \immediate\write16{\fmtname\space version \fmtversion\space
    format preloaded.}%
  \input phystime      
  \input physupdt }    
\immediate\write16{Version \fmtversion\space format loaded.}%


\def\wlog#1{} 
\catcode`\@=11


\def\({\relax\ifmmode[\else$[$\nobreak\hskip.3em\fi}
\def\){\relax\ifmmode]\else\nobreak\hskip.2em$]$\fi}

\def\gappr{\mathpalette\under@rel{>\approx}}
\def\lappr{\mathpalette\under@rel{<\approx}}
\def\gsim{\mathpalette\under@rel{>\sim}}
\def\lsim{\mathpalette\under@rel{<\sim}}
\def\under@rel#1#2{\under@@rel#1#2}

\def\under@@rel#1#2#3{\mathrel{\mathop{#1#2}\limits_{#1#3}}}

\def\under@@rel#1#2#3{\mathrel{\vcenter{\hbox{$%
  \lower3.8pt\hbox{$#1#2$}\atop{\raise1.8pt\hbox{$#1#3$}}%
  $}}}}

\def\widebar#1{\mkern1.5mu\overline{\mkern-1.5mu#1\mkern-1.mu}\mkern1.mu}
\def\parenbar{\mathpalette\p@renb@r}
\def\p@renb@r#1#2{\vbox{%
  \ifx#1\scriptscriptstyle \dimen@.7em\dimen@ii.2em\else
  \ifx#1\scriptstyle \dimen@.8em\dimen@ii.25em\else
  \dimen@1em\dimen@ii.4em\fi\fi \offinterlineskip
  \ialign{\hfill##\hfill\cr
    \vbox{\hrule width\dimen@ii}\cr
    \noalign{\vskip-.3ex}%
    \hbox to\dimen@{$\mathchar300\hfil\mathchar301$}\cr
    \noalign{\vskip-.3ex}%
    $#1#2$\cr}}}


\def\mppae@text{{Max-Planck-Institut f\"ur Physik}}
\def\mppwh@text{{Werner-Heisenberg-Institut}}

\def\mppaddresstext{Postfach 40 12 12, D-8000 M\"unchen 40\else
  P.O.Box 40 12 12, Munich (Fed.^^>Rep.^^>Germany)}

\def\mppaddress{\address{\mppae@text \nl -- \mppwh@text\space --\nl
  \case@language\mppaddresstext}}

\def\mppnum#1{\topright{MPI-Ph/#1}}


\font\fourteenssb=cmssdc10 scaled \magstep2 
\font\seventeenssb=cmssdc10 scaled \magstep3 

\def\letter#1#2{\b@lett@r{26}%
  \centerline{\seventeenssb \uppercase\mppae@text}%
  \centerline{\fourteenssb \uppercase\mppwh@text}%
  \centerline{\strut#1}\vskip.5cm%
  \e@lett@r{\hss\vtop to5cm{\hsize55mm%
    \lftline{\strut}\eightrm  \setbaselineskip=12pt \vfil
    \lftline{F\"OHRINGER RING 6}\lftline{\tenrm D-8000 M\"UNCHEN 40}%
    \lftline{\case@language{TELEFON\else PHONE}: (089) 3 23 08
      \if!#2!\else - #2 \case@language{oder\else or} \fi-1}%
    \lftline{TELEGRAMM:}\lftline{PHYSIKPLANCK M\"UNCHEN}%
    \lftline{TELEX: 5 21 56 19 mppa d}%
    \lftline{TELEFAX: (089) 3 22 67 04}%
    \lftline{POSTFACH 40 12 12}
    \ifx\EARN\undefined\else\vskip5\p@\lftline{EARN/BITNET: \EARN
      @DM0MPI11}\fi \vfil}}}

\def\b@lett@r#1{\endpage \begingroup \doublespace \vglue-#1mm}

\def\e@lett@r#1#2{\skippagenum T\skipheadline T\skipfootline T%
  \line{\vtop to47mm{\lftline{\llap{\vbox to\z@{\vskip171\p@
      \hrule\@width7\p@\vss}\hskip57\p@}\strut}\vskip2mm\vfil
    \addressspacing \dimen@\baselineskip \dimen@ii-2.79ex%
    \advance\dimen@ii\dimen@ \baselineskip\dimen@\@minus\dimen@ii
    \let\nl\cr\use@nl \halign{##\hfil\crcr#2\crcr}\vfil}#1}%
  \vskip1cm\rtline{\thedate}\vskip1cm\@plus1cm\@minus.5cm\endgroup}

\let\addressspacing=\empty


\def\myname{Dr.\ Xxxx Xxxxxxxxxx\nl Physiker}
\def\myaddress{Xxxxxxx Stra\ss e  ??\nl
    \llap{D--8000\quad}M\"unchen ??\nl
    Tel:\ (089) \vtop{\hbox{?? ?? ?? (privat)}%
                      \hbox{3 18 93-??? (B\"uro)}}}
\def\myletter{\b@lett@r{26}\line{\let\nl\cr \use@nl \caps
  \vtop to25mm{\halign{\strut##\hfil\crcr\myname\crcr}\vfil}\hfil
  \vtop to25mm{\tenpoint
    \halign{\strut##\hfil\crcr\myaddress\crcr}\vfil}}%
  \e@lett@r\empty}


\def\firstpageoutput{\physoutput
  \global\output{\setbox\z@\box@cclv \deadcycles\z@}}


\def\veq{\afterassignment\v@eq \dimen@}
\def\v@eq{$$\vcenter to\dimen@{}$$}

\def\veqn{\afterassignment\v@eqn \dimen@}
\def\v@eqn{$$\vcenter to\dimen@{}\eqn$$}

\def\heq{\afterassignment\h@eq \dimen@}
\def\h@eq{$\hbox to\dimen@{}$ }

\def\wlog{\immediate\write\m@ne} 
\catcode`\@=12 

\def\wlog#1{} 
\catcode`\@=11

\outer\def\pthnum#1{\errmessage{***** \string\pthnum\space is no longer
         supported, use \string\mppnum\space instead}}

\def\wlog{\immediate\write\m@ne} 
\catcode`\@=12 



  \overfullrule=0pt
  \parindent=0pt
  \equfull
  \footpar \refsqb
  \crossrestore{strga}
\def\l{\lambda}
\def\a{\alpha}
\def\b{\beta}
\def\d{\delta}
\def\k{\kappa}

\def\p{\partial}

\def\r{\rho}

\def\Ga{\Gamma}
\def\Gacl{\Gamma_{cl}}
\def\ha{{1\over 2}}

\def\N{{\cal N}}

\def\R{{\cal R}}
\def\ga{\gamma}

\def \footstyle{\tenpoint}
\def \G{\Gamma}
\def \Q#1#2 {Q(#1,#2)}

\def \frac#1#2 {\hbox{${#1\over #2}$}}

\def \kdk {\k \p _\k}
\def\mdm {m \p _m}


\RF\JB1{{\caps F. Jegerlehner},
       {Renormalizing the standard model,
        in: Proceedings of the 1990 Theoretical Advanced Study Institute in
        Elementary Particle Physics, Boulder, Colorado, ed. M. Cvetic and P.
        Langacker, 1991, Singapore.}}
\RF\CaSy{{\caps K. Symanzik},
         {\sl Comm. Math. Phys.} {\bf 18} (1970) 227; \lb
         {\caps K. Symanzik}, {\sl Comm. Math. Phys.} {\bf 23} (1971) 49;\lb
    {\caps C.G. Callan}, 
        {\sl Phys. Rev.} {\bf D2} (1970) 1541.}
\RF\R12{{\caps W. Zimmermann},
       {\sl Comm. Math. Phys.} {\bf 15} (1969) 208.}
\RF\R13{
        {\caps J.H. Lowenstein, W. Zimmermann},
       {\sl Comm. Math. Phys.} {\bf 44} \hbox{(1975) 73};
       {\caps J.H. Lowenstein},
       {\sl Comm. Math. Phys.} {\bf 47} (1976) 53.}
\RF\ZiRG{{\caps W. Zimmermann},
          {\sl  Comm. Math. Phys.} {\bf 76} (1980) 39.}
\RF\Bogo{{\caps N.N. Bogoliubov, D.V. Shirkov},
      Introduction to the theory of quantized fields; J.Wiley (1980).}
\RF\LowRG{{\caps J.H. Lowenstein},
                {\sl Comm. Math. Phys.} {\bf 24} (1971) 1.}
\RF\ColRen{{\caps J. Collins}, Renormalization, Cambridge University
             Press 1984.}
\RF\Gross{{\caps D. Gross}, Applications of the renormalization group,
         in: Methods in field theory, Les Houches 1975; ed. R.Balian and
           J.Zinn-Justin, North-Holland 1976.}
\RF\BreiMai{{\caps P.~Breitenlohner, D.~Maison},
              {\sl Comm.~Math.~Phys.} {\bf 52} (1977) 11.}
\RF\CouHil{{\caps R.~Courant, D.~Hilbert}, Methoden  der Mathematischen
            Physik II, Springer Verl.~Berlin 1968; \lb
             {\caps E.~Kamke}, Differentialgleichungen
             II, Akad.~Verlagsgesellsch.~Leipzig 1959. }
\RF\CalSym{{\caps E.~Kraus},
              {\sl Z.~Phys.~C} {\bf 60} (1993) 741 - 750;\lb
            {\caps E.~Kraus},
            The structure of the
            invariant charge in  massive theories with one
       coupling;
      Bern Preprint BUTP-93/26.}
\RF\KrAs{{\caps E.~Kraus},
          Asymptotic normalization conditions and mass independent
          renormalization group functions;\lb Bern Preprint BUTP-94/6}
\RF\GMLRG{{\caps E.C.G.~Stueckelberg, A.~Peterman},
         {\sl Helv.~Phys.~Acta} {\bf 26} \hbox{(1953) 433};
          {\caps M.~Gell-Mann, F.E.~Low},
         {\sl Phys.~Rev.} {\bf 95} (1954) 1300.}
\RF\MarIm{{\caps W.J.~Marciano},
             \PRD20(1979)274* .}
\RF\IZU{{\caps C.~Itzykson, J.-B.~Zuber},
         Quantum field theory, McGraw-Hill 1985}
\RF\BePu{{\caps D.~Bessis, M.~Pusterla},
          \journal{Nuov. Cim} 54 (1968) 234*.}
\RF\ho{{\caps G.~'t Hooft},
       {\sl Nucl.Phys.} {\bf B 61} (1973) 455;\lb
       {\caps J.C. Collins},
        \PRD10(1974)1213* .}

\topright{BUTP-94/6}
\pubdate{April 94}
\title{Asymptotic normalization properties and \nl mass independent
            renormalization group functions}
\author{Elisabeth Kraus
        \rm\footnote
        *{Supported by Deutsche Forschungsgemeinschaft and partially by
            Schweizerischer Nationalfonds
          }}
\address{Institut f\"ur Theoretische Physik, Universit\"at Bern\nl
          Sidlerstrasse 5, CH-3012 Bern, Switzerland}
\abstract{
 Mass independent renormalization group functions in massive
theories are related to normalization properties of the Green functions
in the asymptotic region, where mass effects are neglected.
In this special form the renormalization group invariance
is restricted to the asymptotic region and consequently mass effects cannot
be recoverd by a integration of the renormalziation group equation.
It is shown, that mass independence results in the limit of a large
normalization point, whenever
 a Callan-Symanzik equation exists and contains the same differential
operators as the renormalization group equation.
}
\endpage

\chap{Introduction}
In the development of dimensional regularization schemes have been
constructed, e.g.\ the MS-scheme, which have mass independent
$\beta$-functions and anomalous dimensions in the renormalization
group  equation as well as in the  Callan-Symanzik equation
to all orders of perturbation theory \quref{\ho}.

In this paper we will show that mass independent coefficient functions
in the renormalization group equation indicate that one has chosen
thereby implicitly the
normalization for the independent couplings of the theory in the
asymptotic region, where mass effects are negligible. The intrinsic
normalization properties of all mass independent schemes are therefore
asymptotic ones, and equivalent to the massless theory.
The condition for mass independence in the asymptotic region is the
existence
 of a Callan-Symanzik equation of the same form as the renormalization
group, i.e.~they have to contain both the same differential operators
with respect to fields and couplings.

This insight is important for the use of the respective equations.
Mass independent $\beta$-functions in a massive theory express the fact,
that one has restricted the renormalization group transformations
to the asymptotic region.
As a consequence the momentum dependence of the Green functions, which
one derives from an integration of the renormalization
group equation, is just also the asymptotic one i.e., more
specifically, that of the massless theory. Stated differently:
Mass dependence cannot be recovered by asymptotic renormalization group
transformations. In order to get the full massive invariants of the
renormalization group, one has to integrate the renormalization group
equation with the general mass dependent $\beta $-functions.

In section 1 we give the general renormalization group invariant of
order 1-loop in the massive $\phi^4$-theory, by solving the
renormalization group equation with well-defined boundary conditions.
 Section 2 contains the
proof, that asymptotic normalization properties and mass independent
$\beta $-functions and anomalous dimensions are equivalent.
In the last section we discuss the limit of large momenta or
normalization point, and derive  an approximation  for small
deviations from the asymptotic limit.

\chap{Renormalization group solution
       in the massive $\phi^4$-theory}
Starting point for the considerations is the massive $\phi^4$-theory with
the classical action
$$
\Gacl = \int \ha \p \phi \p \phi - \ha m^2 \phi^2  - {\l \over 4!} \phi ^4.
\EQN{\A1}
$$
In perturbation theory the Green functions are defined by
the Gell-Mann Low formula, a suitable subtraction
scheme and normalization conditions according to
$$\eqalign{ & \p_{p^2} \Gamma_2 (p^2)\bigg|_ {p^2 = \kappa ^2} = 1 \qquad
\qquad \qquad
\Gamma _2 (p^2)\bigg| _{p^2 = m^2} = 0 \cr
& \G _4 (p_1, p_2, p_3, p_4) \bigg|_{{ p_i^2 =  \kappa^2 \hfill} \atop
 { p_i p_j = -   {\kappa^2 \over 3}} }
= - \lambda }
\EQN{\A2}
$$
The normalization point for the coupling and  the wave function
 is taken to be in the Euclidean region ($\kappa^2 <
0$).

 In perturbation theory the dependence of the Green Functions on the massive
parameters is expressed by two partial differential equations, which can be
rigorously derived to all orders of perturbation theory \quref{\CaSy}: the
Callan-Symanzik (CS) equation
$$\bigl(  m \p _m + \kappa \p  _ \kappa +\b _\l \p _\l -
\ga \N
\bigr) \G (\phi)=
 \a \int  \( -m^2 \phi^2 \)_2 \cdot \G (\phi)
\EQN{\A3}
$$
and the renormalization group (RG) equation
$$
\bigl( \kappa \p_ \kappa +\tilde \b _\l \p _\l -
\tilde \ga \N
\bigr) \G (\phi) = 0 \quad
\hbox{with} \quad \N= \int \phi {\d \over \d \phi}.
\EQN{\A4}
$$
The CS-equation describes the breaking of the dilatational
invariance, if one scales the momenta by a common factor $p_i \rightarrow
\epsilon p_i$.
 The RG equation on the other hand
expresses in differential form     invariance of the Green functions under
the renormalization group.

  At the
Euclidean symmetric momentum
$ p_i ^2 = p^2 , p_i p_j = - {p^2 \over 3} , p^2 < 0 $
 the 4-point     vertex in 1-loop order is given by
$$ \eqalign{
 &\Gamma_4
(\hbox{${p^2\over \k^2}$},\hbox{${m^2 \over p^2}$} , \l)\cr
& =  - \l - \hbox{${1\over 16 \pi ^2} {3 \over 2}$} \l^2 \biggl(\,
 \sqrt { 1- \hbox{${3 m^2 \over p^2}$} } \bigl( \ln \bigl(
\sqrt{1- \hbox{${3 m^2 \over p^2}$}} + 1 \bigr) -  \ln \bigl(
\sqrt{1- \hbox{${3 m^2 \over p^2}$}} - 1 \bigr) \bigr)\cr
&\phantom{ = \l - \hbox{${1\over 16 \pi ^2} {3 \over 2}$} \l^2}-
\sqrt { 1- \hbox{${3 m^2 \over \k^2}$} } \bigl( \ln \bigl(
\sqrt{1- \hbox{${3 m^2 \over \k^2}$}} + 1 \bigr) -  \ln \bigl(
\sqrt{1- \hbox{${3 m^2 \over \k^2}$}} - 1 \bigr) \bigr)\biggr) + O(\l^3)\cr
&\equiv  - \l + \l^2 \bigl(Q^{\scriptscriptstyle (1)}
( \frac {m^2}{p^2} ) - Q^{\scriptscriptstyle (1)}( \frac {m^2}{\k^2} )\bigr)
 +O(\l^3),}
\EQN{\loop1}
$$
It  satisfies the normalization condition \queq{\A2} by construction to all
orders.
With the normalization condition \queq{\A2}
 the 2-point function is zero in 1-loop order and consequently the anomalous
dimensions of the CS- and the RG-equation are vanishing.
$$
\Ga_2 (p^2, m^2 ,\k ^2) = p^2 - m^2 + O (\l ^2)\,\, \Rightarrow\,\, \gamma
^{\scriptscriptstyle (1)} = \tilde \gamma
^{\scriptscriptstyle (1)} =0.
\EQN{\andim}
$$
{}From \queq{\loop1} and \queq{\andim}
the RG-function  $ \tilde \b _\l ^ {\scriptscriptstyle (1)}  $
 is calculated to be
$$
 \tilde \b _\l^{\scriptscriptstyle (1)}  (  \frac {m^2}{\k^2} ) =
 \hbox{${1\over 16 \pi ^2} {3 \over 2}$}
 \l^2 \biggl(
 {3 m^2 \over \k^2 }
{  1 \over {\sqrt { 1- \hbox{${3 m^2 \over \k^2}$} }}}
   \ln {
  \sqrt{1- \hbox{${3 m^2 \over \k^2}$}} + 1 \over  \sqrt{1- \hbox{${ 3 m^2
\over \k^2}$} } - 1}
 + 2 \biggr)
\EQN{\BF1}
$$
Only in the limit
of an asymptotic normalization point
$ {m^2 \over \k^2} \to 0$
the RG-function $ \tilde \b _\l  (  \frac {m^2}{\k^2} ) $ becomes
$\k$-independent and coincides with the
CS-function $\b _\l^{\scriptscriptstyle (1)} $:
$$ \b_\l ^{\scriptscriptstyle (1)}
= \lim _{
\scriptscriptstyle \k^2 \to  - \infty} \tilde \b _\l ^{\scriptscriptstyle (1)}
  (  \frac {m^2}{\k^2} )
 = \frac 3{16\pi^2} \l ^2,
\EQN{\CS1}
$$
This is a feature  of the massless limit we will discuss later on
in detail. In the limit $ \k^2 \to 0 $ the
RG-function vanishes:
$$
\lim_{\k^2 \to 0}  \tilde \b _\l ^{\scriptscriptstyle (1)}
 (  \frac {m^2}{\k^2} ) = 0
\EQN{\A14}
$$

According to \queq{\andim}
 the RG-equation is a homogeneous differential equation  in
 order 1-loop,
which can be solved analytically with standard methods \quref{\CouHil}.
 Introducing the variables
$$
  t= \ln \frac {p^2}{\k^2} \qquad \hbox{and} \qquad u =  \frac {m^2}{p^2}
\EQN{\A15}
$$
the RG-equation reads
$$
\Bigl({\p \over \p t } - \frac 12    \tilde \b ^
{\scriptscriptstyle(1)} _\l (u e^t)\Bigr)
\Gamma  = 0 \EQN{\A16}
$$

The characteristic equations are integrated immediately:
$$\eqalign{
{ d\l \over dt } = - \frac 12
\tilde \b _\l^{\scriptscriptstyle (1)} (u e^t) \quad & \Longrightarrow \quad
  - { 1 \over \l(t,u_o, \l_o)} + {1 \over \l _o} =
 Q^{\scriptscriptstyle (1)} (u_o e ^t)
- Q^{\scriptscriptstyle (1)}(u_o)\cr
{ d u \over dt } = 0 \qquad \qquad \qquad &
 \Longrightarrow \quad u(t,u_o,\l_o) = u_o \cr
{ d \Ga_4 \over dt } = 0 \qquad \qquad \qquad &
 \Longrightarrow \quad \Ga _4 (t,u_o, \l_o) =  - \l_o, }
\EQN{\CHE}
$$
with the function $  Q^{\scriptscriptstyle (1)}(y) $ defined in \queq{\loop1}.
In the integration it is unavoidable to fix the
 integration constants $\l _o$ and \hbox{$u_o = \frac {m^2}{p^2} $} of the
characteristic equations by boundary conditions.
According to our construction  the normalization conditions
\queq{\A2} imposed on  the Green functions are satisfied to all
orders and act here as well-defined boundary
conditions by specifying the 2-dimensional boundary manifold, where the
RG-solution has to start in. The equations \queq{\CHE} can be solved with
respect to $\Ga_4 , \l$ and $u$ in order to calculate the
solution $\widebar \Ga _4 (t,u,\l) $ of the RG-equation:
$$\eqalign{
\widebar \Ga_4(t,u,\l)
 &  =  - { \l \over 1  + \l( Q^{\scriptscriptstyle (1)}(u )
  - Q^{\scriptscriptstyle (1)}(u e^t ))}\cr
                      & = - \sum _{i=0}^\infty \l^{i+1}
                         ( Q^{\scriptscriptstyle (1)}(u e^t )  -
           Q^{\scriptscriptstyle (1)}(u ))^i
}
\EQN{\QQ1}
$$
As one easily verifies
$\widebar \Ga_4(t,u,\l)$ is
a RG invariant for an arbitrary momentum $p^2$ and
normalization point $\kappa ^2$, i.e.\ it satisfies the general multiplication
law of the renormalization group:
$$
\widebar \Ga_4(t + t_1 ,u,\l) = \widebar \Ga_4(t,u,\widebar \Ga_4(t_1,u
  e^t  ,\l))
\EQN{\RGM}
$$
It continues the massive
 1-loop Green function in a unique, RG-invariant way to all
orders of perturbation  theory.

Considering the purely asymptotic situation
 $-p^2 \gg m^2$ and $-\k^2 \gg m^2,$  we
get the limit of the massless theory. The respective invariant charge is
 is often called the ``running coupling''
$\bar \l (t) $
\quref{\Gross}:
$$\eqalign{ \bar \l (t) \equiv
  - \widebar \G_{4, {\scriptscriptstyle \infty}}
(t,0,\l) & =
 {\l \over 1 - \l \frac 1{16 \pi ^2} \frac 32 t
} }
\EQN{\largepk}
$$
This expression one would have  obtained, too, if one had calculated the
 RG-solution
 with mass independent
$\b$-function $\tilde \b ^{\scriptscriptstyle (1)}_\l (0) =
\b ^{\scriptscriptstyle (1)}_\l$ (cf.~\queq{\CS1}) and according to the
derivation we have given here it is obvious that \queq{\largepk} is an
asymptotic result, in the sense that the coupling is thought to be fixed in
the asymptotic region {\it and} the momenta are considered in the asymptotic
region, too, where all mass effects are neglected.
Moreover it can be shown quite generally, that
with mass independent functions $\tilde \b _\l $ one only gets the asymptotic
  invariant charge of  the massless theory.
In order to proceed
 we therefore consider once again the solution of the RG-equation especially
\queq{\CHE}.
As we have pointed out there, it
is absolutely mandatory
 to fix the integration constants  by well-defined boundary
conditions. For this reason  one has to know a
point $\k^2  = p^2 f({m^2\over p^2}) $ to all
orders, where the invariant charge has normalization properties, i.e.\
coincides with the coupling.
For well defined normalization conditions as \queq{\A2} such  boundary
conditions are known from the beginning ($\k^2 = p^2$).
In a scheme without specific normalization conditions
 one has to find  a point $\k^2$ where
the invariant  charge has normalization properties in order
to be able to solve the characteristic equations.
 Examples for such schemes are the MS- and
\hbox{$\widebar {\rm MS}$-scheme} \quref{\ColRen}
 or the BPHZL-scheme with $s-1$ at $s=0$ \quref{\R13}.
To all these three schemes it is common that the $\b $-functions and
anomalous dimensions of the CS-equation and RG-equation are the
same and mass independent. In the next section we will prove  quite
generally that in such
schemes one has normalization properties in the asymptotic region.
$$\eqalign{ &
\lim_{\scriptscriptstyle p^2 \to -\infty}
 \p_{p^2} \Gamma_2 (p^2)\bigg|_ {p^2 = \kappa ^2} = 1 \, (+
\rho_{2,1} \l + \rho_{2,2} \l^2 +...)\cr
&
\lim_{\scriptscriptstyle p^2 \to - \infty}
 \G _4 (p_1, p_2, p_3, p_4) \bigg|_{{ p_i^2 =  \kappa^2 \hfill} \atop
 { p_i p_j = -   {\kappa^2 \over 3}} }
= - \lambda \,(+ \rho_{4,1} \l^2 + \rho _{4,2} \l^3 + ...)}
\EQN{\A2inf}
$$
The expressions in the brackets indicate, that we  have to allow also
 reparametrizations of the coupling with
mass independent
coefficients $\rho_{i,j}$.
As an example we calculate the 1-loop 4-point function \queq{\loop1}
 in the schemes
mentioned above and one can easily verify that they have indeed normalization
properties in the asymptotic region as described by \queq{\A2inf}:
$$\eqalign{
{\rm MS}:\hfill   & \>
\G _4(\hbox{${p^2\over \mu^2}$},\hbox{${m^2 \over p^2}$} , \l)  = - \l + \l^2
\bigl(Q^{\scriptscriptstyle (1)} ( \frac {m^2}{p^2} ) +
\frac 1{16 \pi ^2} \frac 32 \ln ( \frac { \mu^2}{ m^2} )
    + \ln 4 \pi - \ga _E) \cr
\widebar {\rm MS}:\hfill & \>
\G _4(\hbox{${p^2\over \bar \mu^2}$},
\hbox{${m^2 \over p^2}$} , \l)  = - \l + \l^2
\bigl(Q^{\scriptscriptstyle (1)} ( \frac {m^2}{p^2} ) +
\frac 1{16 \pi ^2} \frac 32 \ln ( \frac { \bar \mu^2}{ m^2} ))
\cr
{\rm BPHZL}: & \>
\G _4(\hbox{${p^2\over \k^2}$},\hbox{${m^2 \over p^2}$} , \l)  = - \l + \l^2
\bigl(Q^{\scriptscriptstyle (1)} ( \frac {m^2}{p^2} ) +
 \frac 1{16 \pi ^2} \frac 32 \ln (- \frac {4 \k^2}{3 m^2} )),
}
\EQN{\EXA}
$$
where we denoted the normalization point, i.e.\ the unit mass, according
to the general conventions $\mu$ and $\bar \mu$ respectively.
For the qualitative understanding of the consequences of this innocent looking
fact it is useful to observe that with
 the special normalization properties \queq{\A2inf}
  the boundary manifold of the RG-equation is  1-dimensional
(${m^2 \over p^2} =0 $) \quref{\CouHil}
and one cannot reach the non-asymptotic region
by  a RG-transformation due to the singular character of the boundary
conditions, but one will stay in the 1-dimensional asymptotic region.
The  solutions of the system of  characteristic
equations is no more single valued anymore as required in order
to get the full  mass parameter
dependent solution.  The introduction of an  anomalous mass dimension term
will not change the situation.


\chap{Equivalence of asymptotic normalization properties and \nl
         mass independent
     $\beta$-functions}
In this section we will prove the equivalence of asymptotic normalization
properties and mass independent coefficient functions in the CS- and
RG-equation to all orders of perturbation theory. The proof is quite
general and not restricted to the $\phi ^4$-theory, although we
 take only one coupling  for
reasons of transparency.
For this proof we will not
 specify the mass normalization condition and introduce instead
 a $\b $-function for the mass in the RG equation:
$$
\bigl(  \kappa \p_ \kappa +\tilde \b _\l \p _\l
 + \ga _m m \p _m -
\tilde \ga \N
\bigr) \G (\phi) = 0,
\EQN{\RGAM}
$$
the CS-equation is not changed:
$$\bigl(  m \p _m + \kappa \p  _ \kappa +\b _\l \p _\l -
\gamma \N
\bigr) \G (\phi)=
 \a \int  \( -m^2 \phi^2 \)_2 \cdot \G (\phi)
\EQN{\A3}
$$
 In the first part we  impose
normalization conditions in the asymptotic region and show, that the $\b
$-functions and anomalous dimensions of the CS- and RG-equation are  the same
and mass
independent.  In the second part, we start from the assumption that
we have calculated the Green functions in a scheme with mass independent
coefficient functions in the RG- and CS-equation, and prove that the
Green functions of the 4- point vertex and the residuum of the 2-point
function have normalization properties in the asymptotic region. This means
especially, that the differentiation with respect to the mass of the theory
is soft.

To be specific we assume now to have calculated the Green functions with
the following asymptotic normalization condition:
$$\eqalign{ &
\lim_{\scriptscriptstyle p^2 \to \infty}
\p_{p^2} \Gamma_2 (p^2)\bigg|_ {p^2 = \kappa ^2} = 1 \cr
&  \lim_{\scriptscriptstyle p^2 \to \infty}
\G _4 (p_1, p_2, p_3, p_4) \bigg|_{{ p_i^2 =  \kappa^2 \hfill} \atop
 { p_i p_j = -   {\kappa^2 \over 3}} }
= - \lambda \cr }
\EQN{\A2as}
$$
where $\lim_{\scriptscriptstyle p^2 \to \infty} $
means just $- p ^2 \gg m^2 $.
We subtract the CS-equation and the RG-equation,
$$
(m\p_m +( \b_\l - \tilde  \b_\l ) \p _\l - (\ga - \tilde \ga )\N
 - \ga_m m\p _m)
\Ga = \Delta _m    \cdot \G (\phi)
\EQN{\CSRG1}
$$
and test \queq{\CSRG1}  at the normalization point choosing first $p^2$
in the asymptotic region and setting then $p^2 =\k^2$.
Thereby the right-hand-side, the soft mass insertion,
$\Delta _m =  \a \int \( -m^2 \phi^2 \)_2 $, is vanishing for
the 4-point function $\Ga _4$ and for $\p_{p^2} \Ga _2 $ and we remain with
$$\left.\eqalign{
( \b_\l - \tilde  \b_\l ) - 4 \l (\ga - \tilde \ga ) = & 0\cr
 -2 (\ga - \tilde \ga ) =  & 0 } \right\} \qquad
 \eqalign{\Longrightarrow} \qquad
\eqalign{ \b _\l =  & \tilde  \b_\l \cr
\ga = &\tilde \ga}
\EQN{\EQU}
$$
\queq{\CSRG1} simplifies to
$$(m\p_m
 - \ga_m m\p _m)
\Ga = \Delta _m    \cdot \G,
\EQN{\soft}
$$
expressing  the fact,
 that the differentiation with respect to the mass is soft.
To show that the RG-coefficients are mass independent one has to use
 the consistency equation between the CS- and RG-equation, which has
the general form
$$\eqalign{&
(1-\ga_m) \bigl((\k \p _\k \b_\l) \p _\l -  (\k \p _\k \ga) \N \bigr)\Ga \cr
  =  &\,
 (\tilde \b _\l \p_ \l \b _\l -  \b _\l \p_ \l \tilde \b _\l)\p_\l \Ga
-( \tilde \b _\l \p_ \l \ga - \b _\l \p_ \l \tilde \ga ) \N \Ga \cr
  - &\,(\b_\l \p_ \l \ga _m )m \p _m\Ga
   +   \bigl(  \kappa \p_ \kappa +\tilde \b _\l \p _\l
+ \ga_m m \p _m -
\tilde \ga  \N
\bigr)  \bigr( \Delta _m \cdot \G \bigl). \cr}
\EQN{\CONS}
$$
We insert into the consistency equation the result of \queq{\EQU},
test with respect to the Green functions $\Ga _4 $
and $\p_{p^2} \Ga _2 $ and for asymptotic momenta ($p^2 \ne \k^2 ! $),
 where all soft terms, $\Delta _m $
and according to \queq{\soft} $(\b_\l \p_ \l \ga _m )m \p _m \Ga $,
 are  vanishing. In this case \queq{\CONS} simplifies to
$$\eqalign{
(1-\ga_m) \bigl((\k \p _\k \b_\l) \p _\l -  4 (\k \p _\k \ga)  \bigr)\Ga_4
 {\buildrel - p^2 \gg m^2 \over =  } \, & 0 \cr
(1-\ga_m) \bigl((\k \p _\k \b_\l) \p _\l -  2 (\k \p _\k \ga)  \bigr)
\p _{p^2} \Ga_2
{\buildrel - p^2 \gg m^2 \over = } \, & 0 }
\EQN{\CONS2}
$$
which leads order by order to the result
$$ \k \p _\k \b _\l = 0 \qquad \hbox{and} \qquad  \k \p _\k \ga = 0.
\EQN{\MIND}
$$
\queq{\EQU} and \queq{\MIND}  finish the first part of the proof.

To prove the opposite direction we start from the
assumption to  have a scheme where the $\b$-functions
and anomalous dimensions  are  mass-independent
to all orders. Concerning equivalence of the CS-functions and
RG-functions we have only to require that the $\b $-functions agree in
their lowest non-vanishing order:
$$\eqalign{
\hbox{1)} \quad  & \k \p _\k  \b_\l = \k \p _\k  \tilde  \b_\l = 0 \cr
\hbox{2)} \quad & \b_\l^{\scriptscriptstyle (1) } = \tilde
              \b_\l^{\scriptscriptstyle (1) } \cr }
\qquad
\eqalign{
 & \kdk \ga = \kdk \tilde \ga =  0
\cr
 & }
\EQN{\ASS2}
$$
Again in  the RG-equation we allow the appearance of a $\b $-function
for the mass term, as it is common to the generally used mass independent
schemes. We wish to show that asymptotic normalization properties
are  a consequence.
The consistency equation \queq{\CONS} with the assumption \queq{\ASS2 (1)}
inserted reads now:
$$\eqalign{
0  =  &\,
 (\tilde \b _\l \p_ \l \b _\l -  \b _\l \p_ \l \tilde \b _\l)\p_\l \Ga
-( \tilde \b _\l \p_ \l \ga - \b _\l \p_ \l \tilde \ga ) \N \Ga \cr
  - &\,(\b_\l \p_ \l \ga _m )m \p _m\Ga
   +   \bigl(  \kappa \p_ \kappa +\tilde \b _\l \p _\l
+ \ga_m m \p _m -
\tilde \ga  \N
\bigr)  \bigr( \Delta _m \cdot \G \bigl). \cr}
\EQN{\CONSki}
$$
We apply it in 2-loop order  on $ \p _{p^2} \G_ 2 $,
take the limit to  an asymptotic momentum, where the
soft insertion vanishes, and get:
$$
\ga ^{\scriptscriptstyle (1)} = \tilde \ga ^{\scriptscriptstyle (1)}
\EQN{\andim1}
$$
Subtracting now CS- and RG-equation, i.e.\ \queq{\A3} and \queq{\RGAM},
 and  using thereby \queq{\ASS2} and \queq{\andim1}
we get in 1-loop order that the the differentiation
with respect to the mass is soft:
$$ \Bigl( (m\p_m
 - \ga_m m\p _m)
\Ga \Bigr) ^{\scriptscriptstyle (1)}  = \Bigl(\Delta _m    \cdot \G \Bigr)
 ^{\scriptscriptstyle (1)},
\EQN{\soft2}
$$
Specifically this means:
$$ \eqalign{
\mdm \G  ^{\scriptscriptstyle (1)}  _4(p_i,\k^2,m^2) = &\, \bigl( \Delta_m
\cdot \G \bigr)_4  ^{\scriptscriptstyle (1)} (p_i, \k^2, m^2  )
\cr \mdm \p_{p^2} \G  ^{\scriptscriptstyle (1)} _2 (p^2, m^2, \k^2) = &\,
 \p_{p^2} \bigl( \Delta_m
\cdot \G \bigr)_2  ^{\scriptscriptstyle (1)} (p ^2, \k^2, m^2  ).}
\SUBEQNBEGIN{\soft2sa}
$$
Tested at asymptotic symmetric momentum $- p^2 \gg m^2 $ the soft insertions
vanish, showing that the asymptotic expressions only depend on the
ratio $ \frac {p^2}{\k^2} $:
$$\eqalign{
\G_4 ^{\scriptscriptstyle (1)} (p^2, \k^2, m^2,\l)  \>
 {\buildrel  - p^2 \gg m^2
\over \simeq} &\> \l^2\, \G _{4, {\rm as}}  ^{\scriptscriptstyle (1)}
(\frac {p^2}{\k^2} ) \cr
 \p_{p^2} \G  ^{\scriptscriptstyle (1)} _2  (p^2, \k^2, m^2,\l)  \>
 {\buildrel  - p^2 \gg m^2
\over \simeq} & \> \l \,  \omega  ^{\scriptscriptstyle (1)} _ {2, {\rm as}}
 (\frac {p^2}{\k^2} ) }
\EQN{\G4as1}
$$
 The consistency equation \queq{\CONSki}
of order 3-loop
\queq{\ASS2}  tested in the asymptotic region ( $\mdm \G_4
^{\scriptscriptstyle (1)} $ and
  $\mdm  \p_{p^2} \G  ^{\scriptscriptstyle (1)} _2 $
 is soft according to \queq{\soft2}) gives the equalities:
$$\eqalign{
0   = & 5 \,\b^{\scriptscriptstyle (1)} _\l
 (\tilde \b^{\scriptscriptstyle (2)} _\l
   -  \b^{\scriptscriptstyle (2)} _\l
  ) \cr
0 = & 2  \,
\b^{\scriptscriptstyle (1)} _\l (\tilde \ga^{\scriptscriptstyle (2)} -
\ga^{\scriptscriptstyle (2)}) }
\EQN{\B1B2}
$$
and $\tilde \b^{\scriptscriptstyle (2)} _\l =  \b^{\scriptscriptstyle (2)} _\l
$ and $\tilde \ga^{\scriptscriptstyle (2)} =
\ga^{\scriptscriptstyle (2)} $ follows immediately.
Subtracting the CS- and RG-equation in 2-loop order we find the same
equation as above in 1-loop order
$$ \Bigl( (m\p_m
 - \ga_m m\p _m)
\Ga \Bigr) ^{\scriptscriptstyle (2)}  = \Bigl(\Delta _m    \cdot \G \Bigr)
 ^{\scriptscriptstyle (2)}
\EQN{\soft22}
$$
and are able to argue in the same way again (cf.~\queq{{\soft2} -
\shorttag{\G4as1} }):
$$\eqalign{
\G_4 ^{\scriptscriptstyle (2)} (p^2, \k^2, m^2,\l) \>
 {\buildrel  - p^2 \gg m^2
\over \simeq} \> \l ^3 \, \G _{4, {\rm as}}  ^{\scriptscriptstyle (2)}
(\frac {p^2}{\k^2} )\cr
 \p_{p^2} \G  ^{\scriptscriptstyle (2)} _2  (p^2, \k^2, m^2,\l)  \>
 {\buildrel  - p^2 \gg m^2
\over \simeq}  \> \l^2\,  \omega  ^{\scriptscriptstyle (2)} _ {2, {\rm as}}
 (\frac {p^2}{\k^2} ) }
\EQN{\G4as2}
$$
The proof  to all orders is just the same, yielding
 that for asymptotic momenta and mass-independent
$\beta $ functions and anomalous dimensions the variation
with respect to the mass is always soft for the Green functions
in consideration. The asymptotic functions depend therefore
only on the dimensionless ratio $ \frac {p^2}{\k^2} $.
Using now the RG-equation  for the asymptotic Green functions we get the result
that they have  the  following
structure:
$$ \eqalign{
 \G _{4, {\rm as}}  ^{\scriptscriptstyle (n)} = & \l^{n+1} \sum _{i=0 } ^n
\rho_{n,i} \ln ^i (\frac {p^2}{\k^2} )\cr
  \omega  ^{\scriptscriptstyle (n)} _ {2, {\rm as}} = & \l^n
 \sum _{i=0 } ^ {n}
\tilde \rho_{n,i} \ln ^i (\frac {p^2}{\k^2} )
}
\EQN{\lnst}
$$
where $ |p^2| \ge |p_{\infty}^2| \gg m^2. $  With $p_{\infty}^ 2  $ we denote
the smallest momentum from where on one can neglect all mass dependence.
 Taking  $\k^2 $ also in the asymptotic region,
 e.g.~$\k^2 = p_{\infty}^ 2 $
one has the normalization properties we have required in \queq{\A2inf}.

Hence we have shown that mass independence of coefficient functions and
asymptotic normalization are equivalent. This -- we repeat -- means in
particular that we can never recover the non-asymptotic behavior by a
RG-transformation  , but remain ``trapped'' in asymptotics.
This result  is not restricted to the $\phi ^4 $ theory, but
 is valid also for theories with more than one coupling.
 Following the arguments of the proof the most important
ingredient is the existence of the Callan-Symanzik equation with
a truly soft mass-term at the right-hand side and a RG-equation, which
contains -- up to an anomalous mass dimension -- the same differential
operators as the CS-equation.
Therefore the same results hold true for spontaneously broken theories in a
completely unphysical parametrization, i.e.~fixing all the couplings as in
the symmetric theory.
If one chooses in such theories physical normalization conditions
specifying  the masses of the fields, the CS-equation
contains additional differential operators with the consequence, that
the $\beta $-functions depend on the normalization point also in the
asymptotic region \quref{\CalSym}.
Another important input for recovering mass independent $\beta$-functions
is the choice of of a non-exceptional
momentum  as
normalization point. The symmetric point  is non-exceptional and soft terms
 vanish, when it is
driven to infinity.
(This is also the reason, why we have never tested for the full 2-point
function $\G_2 $.)

Therefore it is obvious that -- especially also in theories with spontaneous
 breaking -- mass-independent RG-coefficients are not appropriate to
determine mass effects, because  all
RG-integrations are carried out  in the asymptotic region, where mass
effects are not visible anymore.

\chap{Discussion and applications}

With the general renormalization group invariant
 \queq{\QQ1} one is able to study the limit of large $p^2$
neglecting all terms of order $  {m^2 \over p^2} $ and $  {m^2 \over p^2}
\ln |  {m^2 \over p^2}| $ :
$$\eqalign{
\widebar \G_4 (\hbox{${p^2\over \k^2}$},\hbox{${m^2 \over p^2}$} , \l)
 & {\buildrel  - p^2 \gg m^2
\over \simeq} {- \l \over 1 + \l  -  \bigl(
 \frac 1{16 \pi ^2} \frac 32 \ln (- \frac {4 p^2}{3 m^2} )
 -  Q^{\scriptscriptstyle
(1)}(\hbox{${m^2\over \k^2}$})\bigr) } \cr
 & \; \;\;\> \approx \;\;\; {- \l \over 1 - \l
 \frac 1{16 \pi ^2} \frac 32 \ln (- \frac { p^2}{ m^2} )} }
\EQN{\largep}
$$
where in the last approximation we have assumed that $k$ is somewhere in the
low energy region and
all $\k$-dependent constants multiplied by $\l$ are negligible in comparison to
the logarithmic term. Furthermore  we require to be in
a perturbative domain, i.e.\ $\frac 1{16 \pi ^2} \frac 32
 \l \ln (- \frac {p^2}{m^2} )
  < 1. $
The asymptotic formula
 \queq{\largep} is related to a  region where in a massive  theory  with
positive 1-loop $\b$-function the concept
of improvement works  best: For  a small coupling $\l \ll 1$ fixed  at
a low energy point $\k$, e.g.\ $\k \simeq 0 $, it is then ensured  for
large $-p^2$ that
$$ Q^{\scriptscriptstyle (1)}(\frac {m^2}{\k^2} ) -
  Q^{\scriptscriptstyle (1)}(\frac {m^2}{p^2} ) >1.
\EQN{\IMRE}   $$
As a rough estimate shows, this implies that under these circumstances
 the 1-loop induced invariant charge takes  into account
 the  {\it leading}
terms
in any order
and one is in a situation of summing up  truly
leading logarithms  in the asymptotic
region of the massive
theory. For the electromagnetic coupling,
 where one has a similar situation as in the above one, such
 concepts have been successfully applied in the standard model (see
e.g.~\quref{\JB1}).

On the other hand  choosing the normalization point at large $-\k^2 \gg m^2$,
i.e.\ fixing the coupling  in the asymptotic region, one finds a
corresponding expression:
$$\eqalign{
\widebar \G_4 (\hbox{${p^2\over \k^2}$},\hbox{${m^2 \over p^2}$} , \l)
 & {\buildrel  - \k^2 \gg m^2
\over \simeq} {- \l \over 1 + \l (  Q^{\scriptscriptstyle
(1)}(\hbox{${m^2\over p^2}$} )
+ \frac 1{16 \pi ^2} \frac 32 \ln (- \frac {4 \k^2}{3 m^2} ))
} }
\EQN{\largek}
$$
Due to the alternating sign of the perturbative power series  for
finite $p^2 $ improvement
is now  not naively applicable in the sense of summing up all the leading
contributions, but one rather has to estimate the additional higher order
contributions quite carefully.
(In a UV-asymptotic free theory the situation of \queq{\largep} and
 \queq{\largek}
 is reversed, because the restriction \queq{\IMRE} has the opposite sign.)

Finally we want to use the exact solution of the RG-equation, in order
to study  small deviations from the logarithmic asymptotic behavior
affected by mass terms. This approximation can be derived without taking
care of the sign of the $\beta$-functions and should be relevant especially
in an UV-asymptotically free theory. For these purposes we start from
\queq{\largek}
 with asymptotic normalization conditions and expand $  Q^{\scriptscriptstyle
(1)}(\hbox{${m^2\over p^2}$}) $ for large Euclidean $p^2$:
$$
 Q^{\scriptscriptstyle
(1)}(\hbox{${m^2\over p^2}$} ) =  \frac 12 b_1 \ln | \frac {3 m^2}{4 p^2} |
 + b_{1,1} \frac {m^2}{p^2}  \ln | \frac {m^2}{p^2}  |
+ b_{1,2} \frac {m^2}{p^2}
+ O (\frac {m^4}{p^4}  )
\EQN{\expand}
$$
($b_1$ is determined by the CS-$\beta$-function, $\b _\l ^{\scriptscriptstyle
(1)} = \l^2 b_1 $, i.e.~in the $\phi^4 $-theory
$b_1  = {3 \over 16 \pi ^2}  $.)
The first approximation, where one takes the normalization point
{\it and} the momenta in the asymptotic region, is given in \queq{\largepk}
$$
  \widebar \G_{4, {\scriptscriptstyle \infty}}
(\frac{p^2}{\k ^2}, \frac {m^2}{p^2} , \l)
{\buildrel  - \k^2, p^2 \gg m^2
\over =}
 {\l \over 1 - \l \frac 1{16 \pi ^2} \frac 32 \ln \frac {p^2}{\k ^2}
 } \equiv \bar \l ( \ln \frac {p^2}{\k^2} )
\EQN{\largepk}
$$
In the next step we  want to take into account small deviations
from the asymptotic behavior, i.e.~we have to consider terms of order
$
\bigl( Q^{\scriptscriptstyle
(1)}({m^2\over p^2} )
+  \frac 12 b_1 \ln (- \frac {4p^2}{3m^2} )\bigr) $, but neglect
  all terms of order  $
\bigl( Q^{\scriptscriptstyle
(1)}({m^2\over p^2} )
+  \frac 12 b_1 \ln (-  {4p^2 \over 3m^2}  )\bigr)^2 $.
  Specifically, this means to take into account the terms of order
${m^2 \over p^2  }$  and ${m^2 \over p^2  } \ln |{m^2 \over p^2  } | $
 and to  neglect the terms of order $ {m^4 \over p^4   }$
  for large $p ^2$.
{}From \queq{\largek}  one finds:
$$\eqalign{
 \widebar \G _4 (\hbox{${p^2\over \k^2}$},\hbox{${m^2 \over p^2}$} , \l)
{\buildrel  - \k^2 \gg m^2
\over =}  &
 {- \l \over 1 + \l ( Q^{\scriptscriptstyle (1)}(\hbox{${m^2 \over p^2}$}  )+
  \frac 12 b_1 \ln |\hbox{${4 p^2 \over 3  m^2}$} | )
- \l \frac 12 b_1 \ln  \hbox{${p^2\over \k^2}$ }  } \cr
{\buildrel |\frac{m^4}{p^4} | \ll 1 \over \approx}  &  - \bar \l (t)
+  \bar  \l ^2 (t)
 (Q^{\scriptscriptstyle (1)} (\hbox{${m^2 \over p^2}$}
) + \frac 12 b_1 \ln  | \hbox{${4 p^2 \over 3  m^2}$} |) \cr
= \quad  & - \bar \l (t)
+  \bar  \l ^2 (t) \bigl ( b_{1,1} \frac {m^2}{p^2}  \ln | \frac {m^2}{p^2}  |
+ b_{1,2} \frac {m^2}{p^2} \bigr)
+ O (\frac {m^4}{p^4}  )
}
\EQN{\APPR}
$$
where
$ t = \ln {p^2 \over \k ^2} $. In \queq{\APPR} we have assumed that
 the ``running coupling''  $\bar \l (t) $  is sufficiently small
in the enlarged region of momenta, we consider now.
The  result, we have calculated in a well-defined approximation,
one would have achieved, too, by
solving the RG-equation with a mass-independent
$\b$-function and inserting the solution in the massive  1-loop Green
function. Starting from the complete massive 1-loop invariant charge
the approximation one has performed is obvious, namely one has neglected
 all terms of order
  ${m^4 \over p^4  } $.
 This emphasizes once more that
RG-solutions calculated with mass independent $\b$-functions have
to be used very carefully outside the asymptotic region.

\refout \save{strga} \bye